\documentclass[11pt,onecolumn,a4paper]{article}
\usepackage{amsmath}
\usepackage{amssymb}
\usepackage{graphicx}
\usepackage{bm}
\usepackage{tikz}
\usetikzlibrary{automata, arrows, positioning, calc, bending, decorations.text, decorations.pathreplacing, angles, quotes, fit, patterns}
\usepackage[utf8]{inputenc}
\usepackage{chemformula}
\usepackage{epsfig}
\usepackage{url}
\usepackage{enumitem}
\usepackage{subfig}
\usepackage{epstopdf}
\usepackage{svg}
\usepackage{forest}
\usepackage{algorithm}
\usepackage[noend]{algpseudocode}
\usepackage{booktabs}
\usepackage{pgfplots}
\usepackage{siunitx}
\usepackage{soul}
\usetikzlibrary{decorations.pathreplacing,decorations.markings}
\tikzset{
  font=\normalsize,
  red arrow/.style={
    midway,red,sloped,fill, minimum height=1.5cm, single arrow, single arrow head extend=.6cm, single arrow head indent=.25cm,xscale=0.3,yscale=0.15,
    allow upside down
  },
  black arrow/.style 2 args={-stealth, shorten >=#1, shorten <=#2},
  black arrow/.default={1mm}{1mm},
  tree box/.style={draw, rounded corners, inner sep=.3em},
  node box/.style={white, draw=black, text=black, rectangle, rounded corners},
}

\usepackage{xcolor}
\usepackage{mathtools}

\title{\Large{Leveraging Composition-Based Material Descriptors for Machine Learning Optimization}}
\author{Giovanni Trezza$^{1}$, Eliodoro Chiavazzo$^{1}$\thanks{Corresponding author: eliodoro.chiavazzo@polito.it}\\ \small{\emph{$^{1}$Department of Energy, Politecnico di Torino, C.so Duca degli Abruzzi 24, Torino 10129, Italy}}}
\date{}

\usepackage[left=2cm, right=2cm, top=2cm]{geometry}

\begin{document}

\maketitle

\begin{abstract}

In this study, we evaluate several classifiers and focus on selecting a minimal set of appropriate material features. 
Our objective is to propose and discuss general strategies for reducing the number of descriptors required for material classification. 
The first strategy involves testing whether the critical temperature of the target material property is invariant with respect to binary groups of composition-based features.
We also propose a multi-objective optimization procedure to reduce the set of composition-based material descriptors.
The latter procedure is found to be particularly useful when applied to Bayesian classifiers.
We test the proposed strategies focusing on low-temperature superconductors material data extracted from a public database.
\begin{description}
\item[Keywords] \emph{Machine Learning, Material Classification, Composition-Based Descriptors, Energy Materials, Superconductors}
\end{description}

\end{abstract}

\section{Introduction\label{sec:level1}}
Construction of reliable and predictive models for material properties is becoming an aspect of general interest in a number of areas. 
With a special focus on materials for energy applications, the {\it in silico} prediction of physical properties without resorting to time consuming simulations or expensive experiments is of utmost importance.
One of the reason being that low-cost, long-lasting materials with
high performance is key for energy storage technologies, as it may be responsible from most of the total cost \cite{schmuch2018performance}.

A number of technological areas ranging from the energy up to healthcare sector are being transformed by superconducting materials and may greatly benefit from the discovery of new high performance materials.
Superconductors are materials characterized by zero electrical resistivity when cooled below a superconducting critical temperature $T_{\rm{c}}$ \cite{hirsch2015superconducting}. 
%
%
Due to this fundamental property, such compounds have attracted attention in a wide range of different fields. 
Superconducting Magnetic Energy Storage (SMES) systems allow to store energy by means of a DC current flowing through a superconducting coil; as a consequence, energy can be stored in the resulting magnetic field with almost no loss and can be released back by discharging the coil \cite{johnson2019selecting}. 
Superconducting electromagnets are employed in fusion reactors like tokamak \cite{yuanxi2006first}, Magnetic Resonance Imaging (MRI) \cite{aarnink2012magnetic, hall1991use}, Nuclear Magnetic Resonance (NMR) machines \cite{asayama1996nmr, rigamonti1998basic},  particle accelerators \cite{rossi2012superconducting}. 
Other applications include Superconducting Quantum Interference Devices (SQUIDs) \cite{clarke2004squid}, particle detectors \cite{cristiano2015superconducting}, fast fault current limiters \cite{noe2007high}. 
As such, discovery of new superconductors in the near future is highly desirable and can have a crucial impact on the energy sector (among others). 

Therefore, several recent research studies have made extensive use of Machine Learning (ML)-based approaches.
In particular, Stanev \emph{et al.} \cite{stanev2018machine} trained and validated models both for classification -  prediction of the classes \emph{superconductor}/\emph{non-superconductor} - and for regression - prediction of the critical temperature, employing composition-based features together with the experimental $T_{\rm{c}}$s of known superconductors. 
Konno \emph{et al.} \cite{konno2021deep} represented each chemical formula with four tables, corresponding to the periodic table blocks $s$, $p$, $d$, $f$, with such information being the input of a convolutional Deep Neural Network (DNN) able to predict the critical temperature. 
Le \emph{et al.} \cite{le2020critical} trained and validated a Variational Bayesian Neural Network using superconductors composition-based features for the $T_{\rm{c}}$ prediction. 
Roter \emph{et al.} used only chemical elements and stoichiometry, with no extracted features, to predict the critical temperature \cite{roter2020predicting} and to cluster superconductors \cite{roter2022clustering}.


%
In our work, we deliberately target classical low temperature superconductors, as we could rely upon a high-quality database \cite{supercon}.
%
As such, while collecting the data from the database for training purposes, we discarded all materials containing the following elements: \ch{Fe}, \ch{Ni}, \ch{Cu} (to avoid unconventional superconductivity \cite{stewart2017unconventional}). 
Furthermore, we also removed materials with oxygen to avoid oxides and hopefully include in our analysis materials that are more likely to exhibit ductile behaviour.

After the extraction of 145 composition-based features by means of Matminer \cite{ward2018matminer} for each material formula, we have first trained and validated a tree-based regression model for the prediction of the critical temperature, over which we got insight of the most important features by means of SHAP \cite{lundberg2017unified, lundberg2020local, trezza2022minimal}.
Based on those features, we thus compare several binary classifiers, to distinguish compounds with the critical $T_{\rm{c}}$ exceeding a predefined threshold value from the remaining samples.

\begin{figure}
    \centering
    \includegraphics[width = 0.7\textwidth]{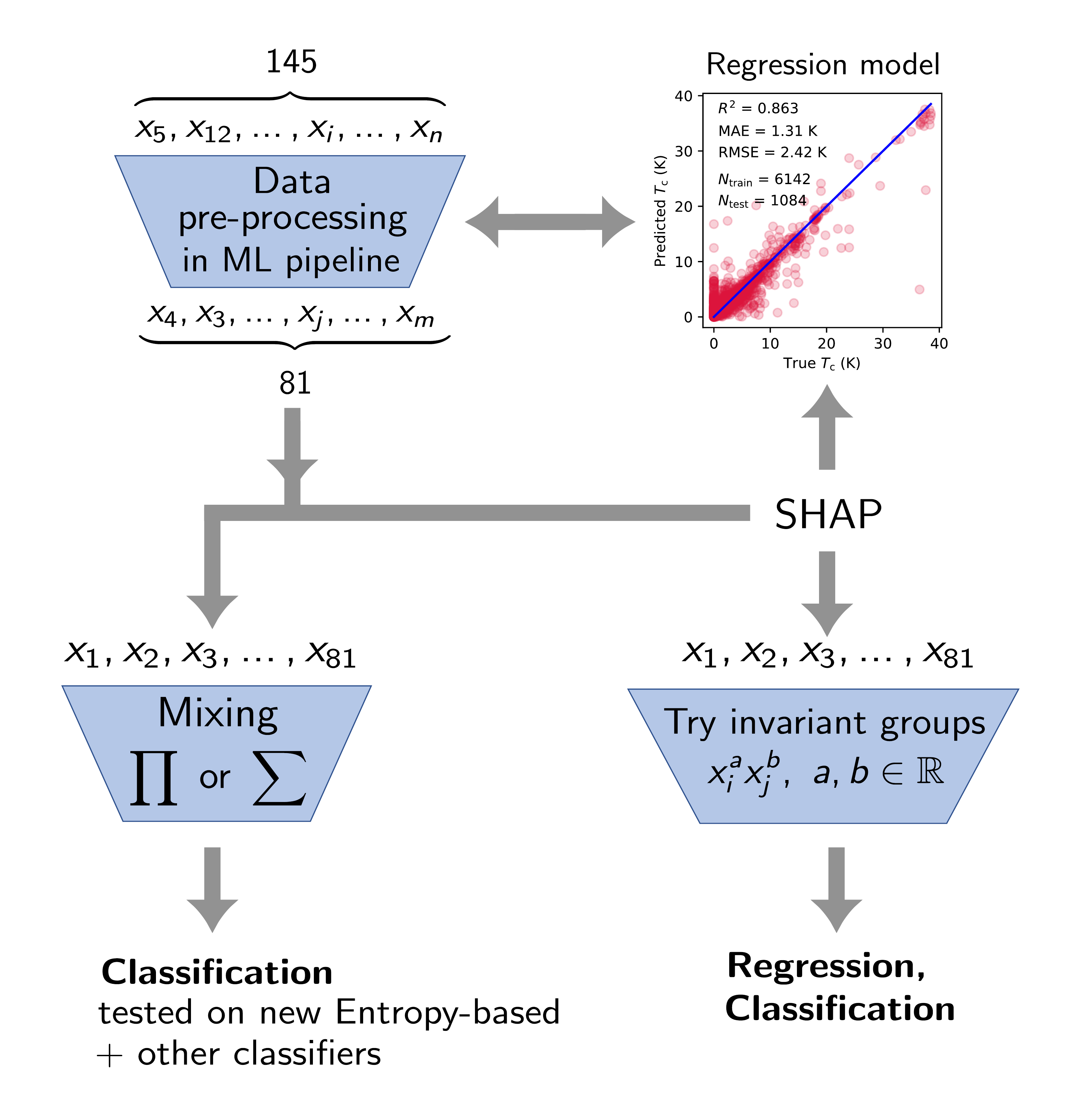}
    \caption{Overview of the protocol used to find a reduced set of ruling descriptors for conventional superconductivity and for the construction of optimized mixed features. Over 7000 chemical compositions have been featurized with 145 descriptors. A regression model has been trained and validated over this dataset, and during the pre-processing routines (i.e., feature reduction by means of linear correlation analysis, descriptors variance analysis, correlation analysis with the $T_{\rm{c}}$, see Supplementary Note 4 for details), many of those features are discarded, ending up with 81 descriptors. 
    By means of SHAP, those 81 features are ranked in terms of importance. The work aimed at finding optimized mixed features for both regression/classification in the form $x_i^ax_j^b$; and for classification, with power or linear combination of the primitive features. The latter descriptors have been tested over both new entropy-based classifiers and other classifiers.}
    \label{fig:visual_abstract}
\end{figure}

A special focus of this study is on the identification and construction of a minimal and optimal set of key material descriptors (or features) to be adopted for classification purposes.
To this end, we pursue two main strategies as briefly described below and schematically represented in Fig.~\ref{fig:visual_abstract}:
\begin{itemize}
    \item We first perform the aforementioned SHAP analysis and establish a descriptor ranking based on the relevance of single features $\{x_{i= 1,...,n} \}$, where $n$ is the maximum number of adopted features;
    \item In the spirit of the work by Tegmark and collaborators \cite{udrescu2020ai}, we propose a general approach for investigating possible symmetries of the target quantity (here $T_{\rm{c}}$) with respect to groups of the originally chosen features (according to the order suggested by SHAP). Without loss of generality, we focus on feature binary groups in the form $x_i^ax_j^b$, with $a, b\in\mathbb{R}$ being properly selected constants. To this end, a proper algorithm based on the computation of the output gradient with respect to the input features by means of a Deep Neural Network (DNN) is discussed.
    \item Finally, a general framework for drastically reducing the number of the classifier features is proposed in the form of both single and multi-objective optimization problem. Among other purposes, the latter approach proves particularly convenient to {\it synthetically} construct new descriptors particularly suited for Bayesian type classifiers, including a novel entropy-based classifier introduced and tested in this work. 
%
%
\end{itemize}


Finally, we use the classifiers with the best performance, to rank $\sim40,000$ compounds from Materials Project \cite{jain2013commentary} and not occurring in the SuperCon.


\section{Methods}

\subsection{Dataset creation}
First, we aim at the construction of a database suitable for ML regression to predict the critical temperature.
In particular, the SuperCon database \cite{supercon} collects both inorganic (under the class ``Oxide and Metallic") and organic materials (under the class ``Organic").
We have considered only the entire subset of inorganic compounds, consisting of $\sim33,000$ entries, of which $\sim7,000$ have no $T_{\rm{c}}$; for those latter compounds, we have assumed $T_{\rm{c}}=0\, \si{\kelvin}$. 
We have thus dropped all materials whose formulae contain symbols like `-', `+', `,', strings like `X', `Z', `z' when not included in meaningful elements symbols (e.g., `Zn'), and with $T_{\rm{c}}>150\, \si{\kelvin}$, resulting in a reduction of the number of compounds to $\sim26,000$.
Furthermore, after normalizing the formulae stoichiometry, we have followed the same approach explained by Stanev \emph{et al.} \cite{stanev2018machine} for dealing with the duplicates. 
In particular, when the same compound was reported with different $T_{\rm{c}}$ values, we have retained it with the average critical temperature only if $\textrm{std}(T_{\rm{c}})\leq5\, \si{\kelvin}$, otherwise we dropped all of its occurrences, ending up with $\sim{16,000}$ unique compounds. 
Moreover, we have taken into account only the classical superconductivity: we have dropped materials with \ch{Ni}, \ch{Fe}, \ch{Cu} and \ch{O} (to avoid oxides). 
We have finally dropped four outliers with $T_c>50\, \si{\kelvin}$ (see Supplementary Note 6 for details). 
These latter steps left $\sim{7200}$ materials for classical superconductivity, of which $\sim{6700}$ have $T_{\rm{c}}<15\, \si{\kelvin}$. 
We have addressed those cleaning pre-processing by employing the Python Pandas package \cite{mckinney2012python}. 

We have then converted each {\it brute} formula into 145 composition-based descriptors by means of Matminer \cite{ward2018matminer}. 
Specifically, as stated by Ward \emph{et al.} \cite{ward2016general}, they include stoichiometric attributes (depending on the elements' ratios), elemental property statistics (representing mean, absolute deviation, minimum and maximum of 22 atomic properties, e.g., atomic number, atomic radii), electronic structure attributes (corresponding to the average fraction of electrons in \emph{s}, \emph{p}, \emph{d}, \emph{f} valence shells over all the elements in the compound) and ionic compound attributes (including whether it is possible to form an ionic compound assuming all elements are present in a single oxidation state).

\subsection{Regression models and descriptors choice}
As a second step, we have trained and validated two different regression ML models for the prediction of the critical temperature. 
The former is a tree-based model, allowing the exact computation of the coefficients of importance in terms of the $T_{\rm{c}}$ by means of the Tree SHAP interpretation algorithm \cite{lundberg2020local, lundberg2017unified}. 
The latter is a Deep Neural Network (DNN), allowing the computation of the gradient of the critical temperature with respect to the input features, namely $\nabla T_{\rm{c}}(x_1, \dots, x_n)$, which is necessary for the identification of the invariant groups in the form $x_i^ax_j^b$. 

Specifically, the former model is an ETR-based pipeline, with hyperparameter tuning in 5-fold cross validation, trained over the 85\% of the dataset and tested over the remaining 15\%.
The latter is a DNN trained and validated over the 85\% of the database - of which the 85\% has been used for the training and the remaining 15\% for the validation - and tested over the remaining 15\%. 
For further details about the regression models, please refer to Supplementary Notes 3 and 4.

\subsection{Invariant groups identification procedure}
For the identification of the invariant binary groups in the form $x_i^ax_j^b$ we have applied the following procedure.

The critical temperature is a function of more variables, namely $T_{\rm{c}} = T_{\rm{c}}(x_1, \dots, x_n)$. 
If $T_{\rm{c}}$ is invariant with respect to a group of features in the form $x_i^ax_j^b$, when this group is a constant $\overline{c}$ -  even varying the components $x_i, x_j$ separately - the critical temperature does not change as well. This yields
\begin{equation}
a\ln(x_i) + b\ln(x_j) = c
\label{eq:const}
\end{equation}
where $c = \ln(\overline{c})$. If $c$ is constant, $\textrm{d}c=0$; so, Eq.~\ref{eq:const} can be rewritten as, 
\begin{equation}
    a\frac{\textrm{d}x_i}{x_i} + b\frac{\textrm{d}x_j}{x_j} = 0.
\end{equation}
An orthogonal vector $\bf{n}$ to the locus of points with $\textrm{d}c=0$ in a point $\overline{\mathbf{x}_0} = (x_{i,0}, x_{j,0})$ has components $(a/x_{i,0}, b/x_{j,0})$, which normalized becomes the following unit vector:
\begin{equation}
\hat{\mathbf{n}} = \left(\frac{a}{x_{i,0}\sqrt{\left(\frac{a}{x_{i,0}}\right)^2 + \left(\frac{b}{x_{j,0}}\right)^2}},
\frac{b}{x_{j,0}\sqrt{\left(\frac{a}{x_{i,0}}\right)^2 + \left(\frac{b}{x_{j,0}}\right)^2}}
\right).
\end{equation}

The condition of invariance with respect to the group $x_i^ax_j^b$ requires that the components of the gradient $\nabla T_{\rm{c}}(x_1, \dots, x_n)$ are aligned with $\hat{\mathbf{n}}$ in $\overline{\mathbf{x}_0}$. 
This yields the system
\begin{equation}
\begin{cases}
    \left(\frac{\partial\widetilde{T_{\rm{c}}}}{\partial x_i}\right)_{\overline{\mathbf{x}_0}}-\hat{n}_1 = 0 \\
    \left(\frac{\partial\widetilde{T_{\rm{c}}}}{\partial x_j}\right)_{\overline{\mathbf{x}_0}}-\hat{n}_2 = 0
\end{cases}
\label{eq:system}
\end{equation}
where 
\begin{equation}
\begin{split}
\left(\frac{\partial\widetilde{T_{\rm{c}}}}{\partial x_i}\right)_{\overline{\mathbf{x}_0}} & = \left(\frac{\partial{T_{\rm{c}}}}{\partial x_i}\right)_{\overline{\mathbf{x}_0}}\left(\left(\frac{\partial{T_{\rm{c}}}}{\partial x_i}\right)_{\overline{\mathbf{x}_0}}^2 + \left(\frac{\partial{T_{\rm{c}}}}{\partial x_j}\right)_{\overline{\mathbf{x}_0}}^2\right)^{-1/2}\\
\left(\frac{\partial\widetilde{T_{\rm{c}}}}{\partial x_j}\right)_{\overline{\mathbf{x}_0}} & = \left(\frac{\partial{T_{\rm{c}}}}{\partial x_j}\right)_{\overline{\mathbf{x}_0}}\left(\left(\frac{\partial{T_{\rm{c}}}}{\partial x_i}\right)_{\overline{\mathbf{x}_0}}^2 + \left(\frac{\partial{T_{\rm{c}}}}{\partial x_j}\right)_{\overline{\mathbf{x}_0}}^2\right)^{-1/2}
\end{split}
\end{equation}
and $n_1, n_2$ represent the two components of the unit vector $\hat{\mathbf{n}}$. If the non-linear system in Eq.~\ref{eq:system} is satisfied for the same exponents $(\overline{a}, \overline{b})$ over all the domains of the variables $(x_i, x_j)$ - namely $x_{i, \textrm{min}}\leq \forall x_i\leq x_{i, \textrm{max}}$ and $x_{j, \textrm{min}}\leq \forall x_j\leq x_{j, \textrm{max}}$ - the group $x_i^{\overline{a}}x_j^{\overline{b}}$ is an intrinsic variable.

From the practical viewpoint, this has required the computation of the gradient $\nabla T_{\rm{c}}(x_1, \dots, x_n)$, where the function $T_{\rm{c}}(x_1, \dots, x_n)$ is represented by the DNN - built by means of Tensorflow \cite{tensorflow2015-whitepaper} - linking the critical temperature with the input features. 
In particular, once the network has been trained and validated, we have employed the automatic differentiation to compute those partial derivatives. 
Specifically, for getting e.g., $\left(\frac{\partial T_{\rm{c}}}{\partial x_j}\right)$ over all the domain of the variable $x_j$, we have fixed all the other variables $(x_1, \dots, x_{j-1}, x_{j+1}, \dots, x_n)$ to their average values in the original database. Finally, for each group of two features $x_i$ and $x_j$, we have computed 100 different times the values of $a$ and $b$ respectively, comparing them for getting insight of possible invariance.
The above approach has been tested in the Supplementary Note 2 by means of properly designed synthetic example.

\subsection{QEG-based probabilistic classifier}

In addition to more classical classifiers, we have attempted the construction of maximum Shannon entropy-based probabilistic classifier based on the concept of Quasi Equilibrium Manifold as defined in \cite{gorban2005invariant,gorban2018model} and implemented in the discrete form of Quasi Equilibrium Grid (QEG) as discussed in \cite{chiavazzo2009invariant,chiavazzo2008quasi,chiavazzo2011adaptive,chiavazzo2012approximation}. 
The main idea is described below.
Given a number $s$ of important descriptors, we have first discretized those features from the original dataset by means of a $s\rm{-dimensional}$ binning, where each descriptor accounts for a number of bins $N_1, \dots, N_s$. 
Our aim was thus to build a probability distribution $p(x_1, \dots, x_s)$ having the same mean vector and covariance matrix of the original binned data; among the infinite distributions respecting those bounds, we were interested in the one maximizing the Shannon Entropy. 
Given the total number of $s-\rm{dimensional}$ bins $N = N_1\times\dots\times N_s$, the general idea consists in starting with a flattened probability distribution $\mathbf{p}^0=(p_1, \dots, p_{N})^0$ and ending up with a corrected distribution, which respects the constraints of mean vector and covariance we have imposed. 
The QEG guarantees that, if $\mathbf{p}^0$ lies on the surface of maximum Shannon Entropy, also any corrected distribution will lie on the same surface. 
For this reason, we have always chosen $\mathbf{p}^0$ as the uniform distribution, where each entry is $1/N$.

To this end, we have defined a matrix $\mathbf{m}\in\mathbb{R}^{l\times N}$, where $l = (3s + s^2)/2$. 
The first $s$ rows of $\mathbf{m}$ represent the binning of those $s$ descriptors. 
The remaining $l-s$ rows represent the covariance matrix entries of those $s$ descriptors; namely, given the integers $i, j\in[1, s]$, with $i\geq j$, the generic row of $\mathbf{m}$ among the last $l-s$ rows is the result of the element-wise product $(\mathbf{m}_i-\mu_i)(\mathbf{m}_j-\mu_j)$, where $\mu_i, \mu_j$ are the means of the $i\rm{-th}$ and $j\rm{-th}$ descriptor respectively, while $\mathbf{m}_i, \mathbf{m}_j$ are the $i\rm{-th}$ and $j\rm{-th}$ rows of $\mathbf{m}$ respectively.
Furthermore, we define the matrix $\mathbf{E} = (\mathbf{m}, \mathbf{1})^\top$, where $\mathbf{1}=(1, \dots, 1)\in\mathbb{R}^{N}$ represents the normalization condition for probability. Let denote the null space of $\mathbf{1}$ with $\bm{\rho}\in\mathbb{R}^{N\times(N-1)}$ and the null space of $\mathbf{E}$ with $\mathbf{t}\in\mathbb{R}^{N\times(N-l-1)}$. 
We thus construct a square matrix $\mathbf{A}\in\mathbb{R}^{(N-1)\times (N-1)}$ and a vector $\mathbf{b}\in\mathbb{R}^{N-1}$. For the first $N-l-1$ rows, the generic elements correspond to

\begin{equation}
\begin{split}
    A_{ij} & =  \left\langle\mathbf{t}_i, \langle\textrm{diag}(-1/\mathbf{p}), \bm{\rho}_j)\rangle\right\rangle  \\
    b_i & =  \langle\left(1+\ln(\mathbf{p})\right), \mathbf{t}_i\rangle 
\end{split}
\end{equation}

while for the remaining $l$ rows they are

\begin{equation}
\begin{split}
    A_{ij} & =  \left\langle\bm{\rho}_j, \mathbf{m}_i\right\rangle  \\
    b_i & =  0
\end{split}
\end{equation}

where $\mathbf{t}_i$, $\bm{\rho}_j$ and $\mathbf{m}_i$ are the $i\textrm{-th}$ column of $\mathbf{t}$, the $j\textrm{-th}$ column of $\bm{\rho}$ and the $i\textrm{-th}$ row of $\mathbf{m}$ respectively, $\mathbf{p}$ represents the flattened probability distribution at the current iteration step, $\langle\cdot,\cdot\rangle$ denotes the dot product.

 The correction procedure for the $i\textrm{-th}$ bound is carried out as follows:
(i) the starting point is computed as $\langle\mathbf{m}_i, \mathbf{p}\rangle$,
(ii) the desired value is computed as $\langle\mathbf{m}_i, \widetilde{\mathbf{p}}\rangle$, where the $j\textrm{-th}$ entry of $\widetilde{\mathbf{p}}$ is the number of items belonging to the $j\textrm{-th}$ $s\textrm{-dimensional}$ bin over the total number of items (namely, the frequency),
(iii) the resulting residual is filled by solving the system $\mathbf{A}^k\mathbf{p}^{k+1}=\mathbf{b}^k$ iteratively, by replacing time by time $b_{N-l-1+i}$ with a correction step $\varepsilon$, where $\mathbf{p}^{k+1} = \mathbf{p}^k + \delta\mathbf{p}^{k}$ and $\delta\mathbf{p}^{k}$ represents the correction resulting from the $k\textrm{-th}$ iteration, (iv)  when the correction over the $i\textrm{-th}$ bound is complete, the correction over the $i+1\textrm{-th}$ bound can start, by imposing $b_{N-l-1+i} = 0$ and $b_{N-l-1+i + 1} = \varepsilon$.

\section{Results and discussion}
As mentioned above, and in line with others in the literature \cite{stanev2018machine, konno2021deep, le2020critical, roter2020predicting, roter2022clustering}, 
we adopted a convenient source of data, namely the SuperCon database \cite{supercon} which collects the values of critical temperatures $T_{\rm{c}}$ for superconducting materials known from literature. 
To our knowledge, SuperCon represents the largest database of its kind, from which we have extracted a list of $\sim16,000$ materials. 
Beyond the $T_{\rm{c}}$ values, the SuperCon database provides only the chemical composition of a compound. 
The latter info was thus converted into meaningful features by means of Matminer \cite{ward2018matminer}, allowing us to associate the normalized brute formula of each compound with 145 composition-based descriptors (see Methods for further details).

Armed with such features, we can thus make use and compare the performance of several classifiers aiming at predicting the probability for a compound to be a superconductor candidate.
In our study, we made use of known classifiers.
In addition, we also investigate a Bayesian type classifier based on the concept of Quasi-Equilibrium Manifold \cite{gorban2005invariant, gorban2018model,chiavazzo2009invariant,chiavazzo2011adaptive}, as detailed above.

\subsection{Models for predicting the critical temperature value}
%
%
We have first trained and validated an Extra Trees Regressor (ETR)-based pipeline with hyperparameter tuning in 5-fold cross-validation (see Supplementary Notes 4 and 7 for details). 
By means of the Tree SHAP algorithm \cite{lundberg2020local, lundberg2017unified}, we have sorted the input features in terms of their relevance with respect to the prediction of the $T_{\textrm{c}}$.
\begin{figure}
    \centering
    \includegraphics[width = 0.8\textwidth]{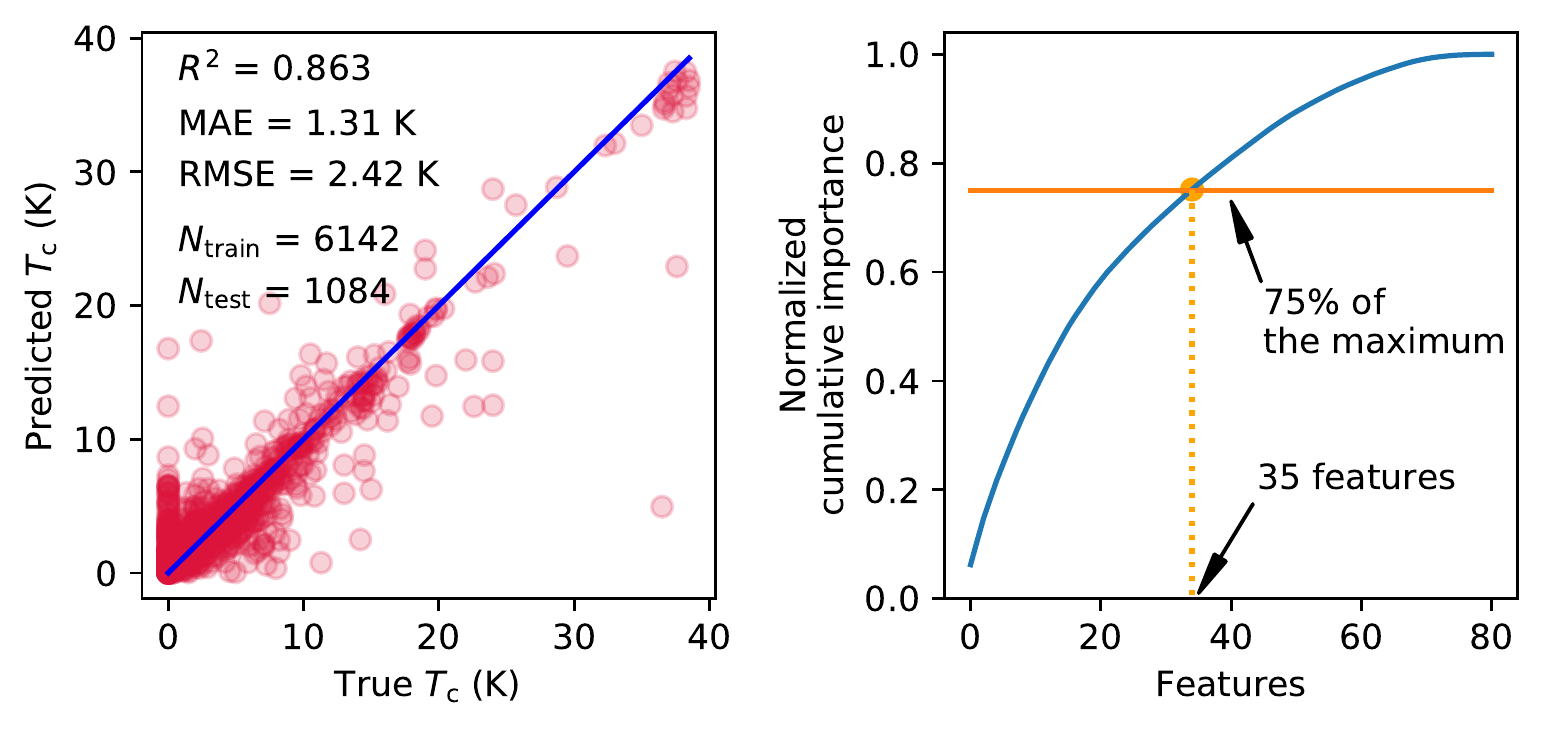}
    \caption{Predictions and corresponding normalized cumulative curve for the coefficients of importance of the ETR model. Model performances are shown in terms of coefficient of determination $R^2$, mean absolute error (MAE), and root mean squared error (RMSE), with the size of training and testing sets $N_{\rm{train}}$ and $N_{\rm{test}}$, respectively.}
    \label{fig:model}
\end{figure}
Model performances with the corresponding cumulative importance curves of the ruling descriptors are reported in Fig.~\ref{fig:model}. 
During the data preprocessing routines, the trained pipeline (i.e., feature reduction by means of linear correlation analysis, descriptors variance analysis, correlation analysis with the $T_{\rm{c}}$ and ML with hyperparameter tuning, see Supplementary Note 4 for details) already drops a significant number of the 145 features, thus confirming that many of the initially selected descriptors do not significantly affect the chosen target property.
%
%
In particular, the final model only includes 81 descriptors.

\begin{figure}
    \centering
    \includegraphics[width = 0.7\textwidth]{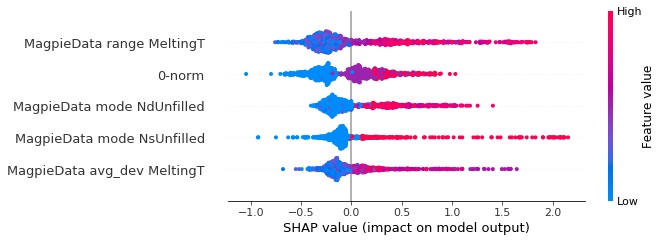}
    \caption{The five most important features according to SHAP ranking for $T_{\rm{c}}$. For each feature (i.e., each line), 1084 dots are shown, representing the entire testing sets used for computing the related SHAP values (impacts on the model output, horizontal axes); the color represents the corresponding feature value, the features are sorted according to the mean over the absolute SHAP values.}
    \label{fig:ETR-metallic}
\end{figure}

Importantly, Fig.~\ref{fig:ETR-metallic} shows the SHAP rankings of the five most meaningful descriptors for the aforementioned model. 
Table \ref{tab:descriptors} summarizes the physicochemical meaning of the identified descriptors, based on the complete list by Ward \emph{et al.} \cite{ward2016general}. 
The entire list of variables, together with their cumulative importance, the trained models, and the datasets on which they have been trained are publicly available online (see Data availability and Code availability).
\begin{table*}[h]
\centering
\caption{\label{tab1}Relevant composition-based descriptors and their meaning \cite{ward2016general}.}
\begin{tabular}{ll}
\toprule
\textbf{Descriptor name} & \textbf{Meaning}\\
\midrule
MagpieData range MeltingT & Range of melting $T$ over the elements of a compound\\
0-norm & Number of different chemical species\\
MagpieData mode NdUnfilled & Mode of $d$ unfilled orbitals over the elements\\
MagpieData mode NsUnfilled & Mode of $s$ unfilled orbitals over the elements\\
MagpieData avg\_dev MeltingT & Average absolute deviation of melting $T$ over the elements\\
\bottomrule
\end{tabular}
\label{tab:descriptors}
\end{table*}

\subsection{Invariant groups}
In a first attempt of reducing the number of input features within the above models, we decided to investigate on the possible existence of symmetries of the obtained regression models.
In particular, we have been focusing on the possible invariance of the target property (here the critical temperature) with respect to binary groups of the form: $x_i^ax_j^b$.
In this study, we restrict to binary groups, although we are confident that the approach can be also generalized to groups concurrently involving a larger number of features. 
To this end, as discussed in the Methods section below, it is necessary to get access to the gradient of the critical temperature with respect to the input features, namely $\nabla T_{\textrm{c}}(x_1, \dots, x_n)$.
We have thus approximated the function $T_{\textrm{c}}(x_1, \dots, x_n)$ with a Deep Neural Network (DNN), which is - to our knowledge - a convenient model allowing to compute that gradient by means of automatic differentiation.

\begin{figure}
    \centering
    \includegraphics[width = 0.8\textwidth]{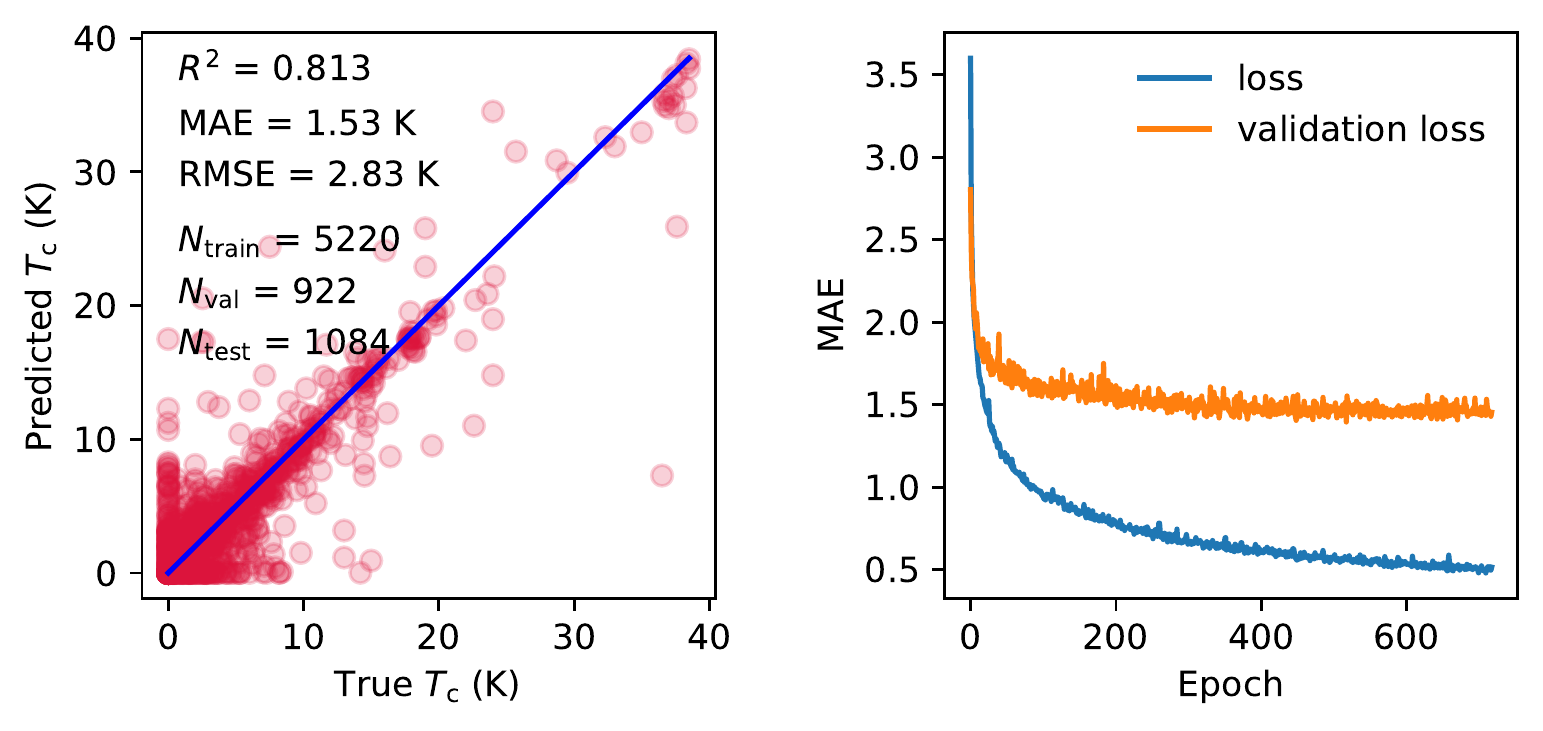}
    \caption{Predictions over the testing set and corresponding loss curves for the DNN regression model. Model performances are shown in terms of coefficient of determination $R^2$, mean absolute error (MAE), and root mean squared error (RMSE), with the sizes of the training, the validation and the testing sets, $N_{\textrm{train}}$, $N_{\textrm{val}}$, $N_{\textrm{test}}$ respectively.}
    \label{fig:DNN-metallic-ML}
\end{figure}

As input features of the DNN, we have employed the same 81 relevant descriptors of the above ETR-based pipeline. 
We have thus splitted the dataset into three parts: (i) a training set, (ii) a validation set to get insight of possible overfitting, (iii) a testing set to effectively evaluate the model performances.
Fig.~\ref{fig:DNN-metallic-ML} shows the predictions over the testing set, together with the model performances and the corresponding loss with respect to the number of epochs. 
Specifically, no overfitting is found. 
More details about the DNN structure are shown in the Supplementary Note 3. 
We have thus looked for possible invariant groups in the form $x_i^ax_j^b$ among the 45 different combinations of the most relevant 10 features according to the SHAP-based ranking above. 
On the basis of our investigations, we can conclude that the critical temperature of the examined materials presents no invariance with respect to the tested binary groups.

\subsection{Entropy-based binary classifiers}
In this section, we introduce and test a special Bayesian type classifier as detailed below.
We have considered the first two features of the SHAP ranking for constructing a Shannon Entropy-based probabilistic classifier according to the Quasi Equilibrium Grid (QEG)-based procedure reported above in the Methods.
\begin{figure}
    \centering
    \includegraphics[width = 0.4\textwidth]{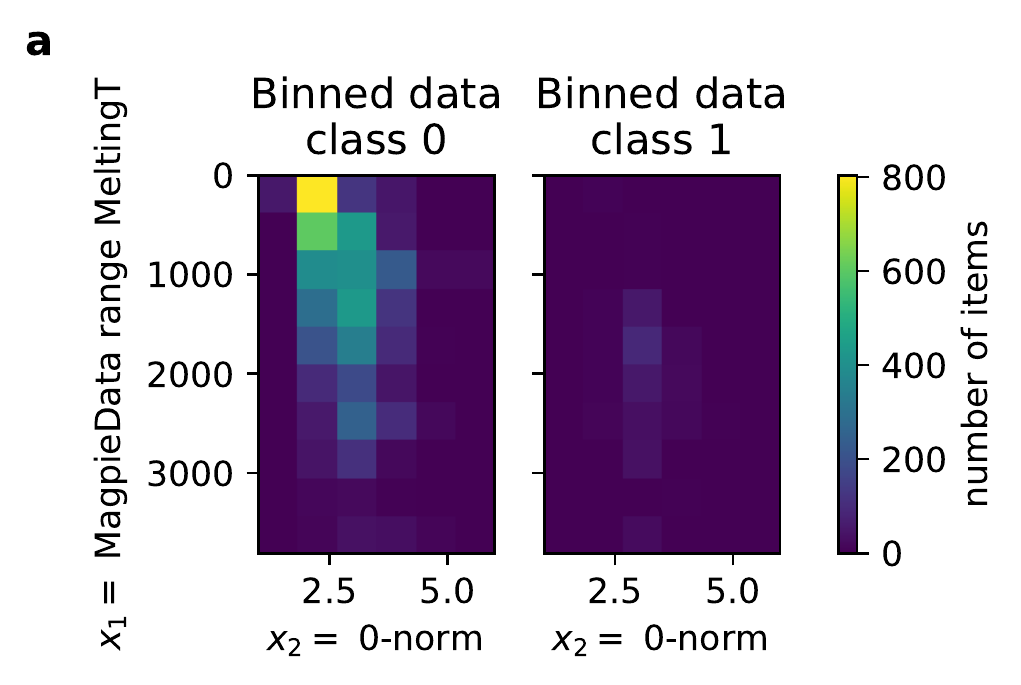}
    \includegraphics[width = 0.4\textwidth]{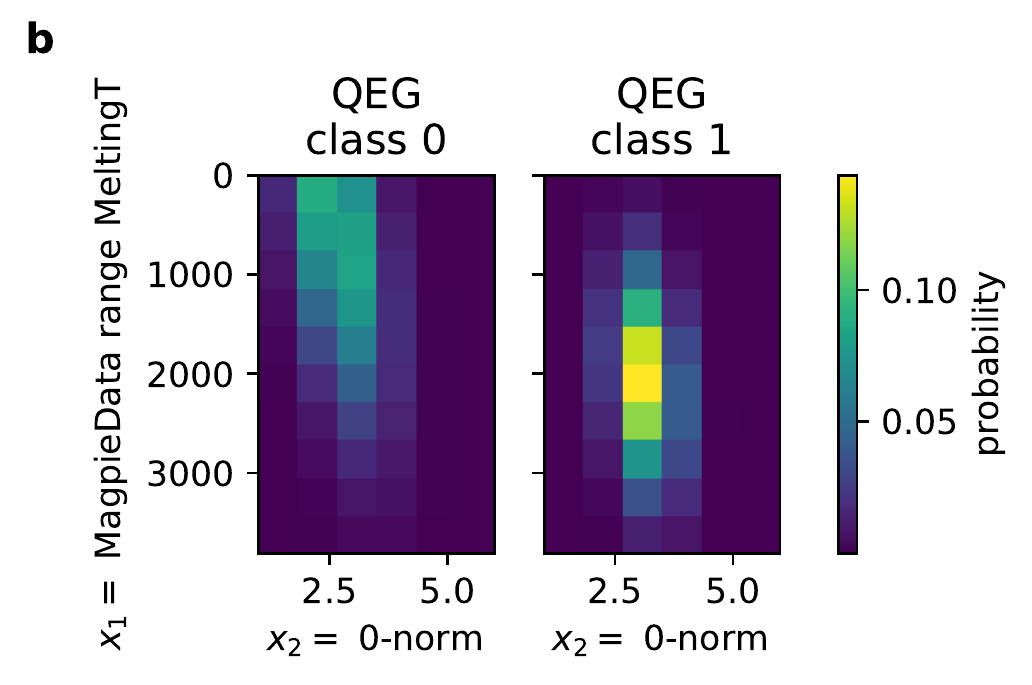}
    \includegraphics[width = 0.4\textwidth]{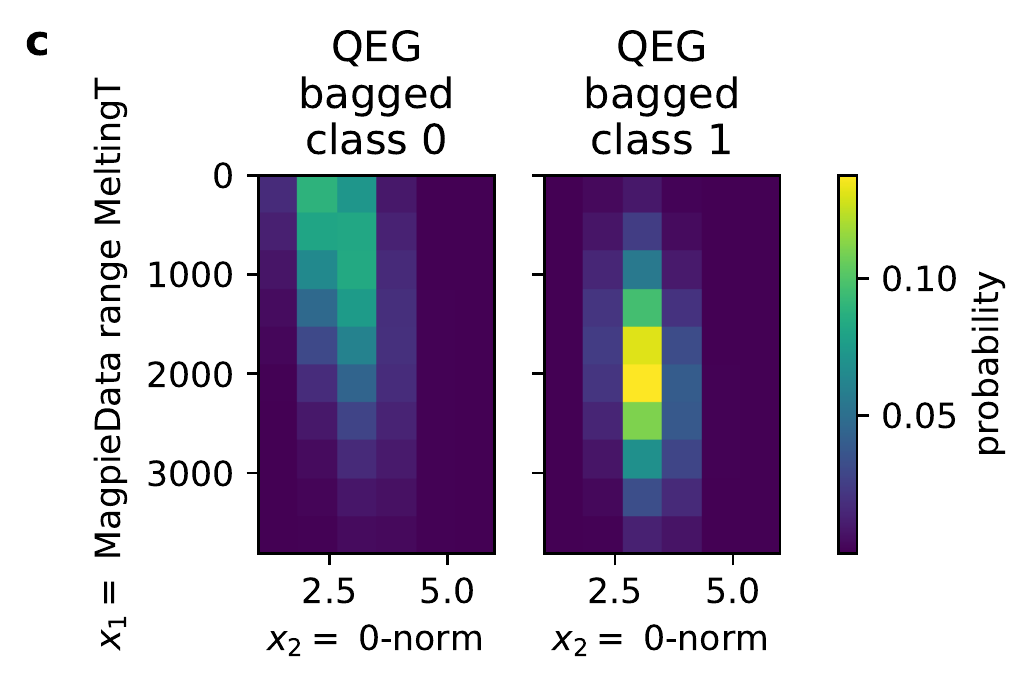}
    \caption{Probabilistic classifier. \textbf{a} 2-dimensional binning, with 10 bins for the first variable and 5 bins for the second, of the two most relevant features $x_1, x_2$ according to the SHAP ranking for superconductors showing $T_{\rm{c}}<15\, \si{\kelvin}$ and $T_{\rm{c}}\geq15\, \si{\kelvin}$ respectively among the training set (namely, 85\% of materials); \textbf{b} QEG solution of corresponding maximum Shannon entropy probability distribution; \textbf{c} QEG solution of corresponding maximum Shannon entropy probability distribution, bagged case.}
    \label{fig:qeg-metallic}
\end{figure}
In particular, we have binned those two features separately for superconductors with both $T_{\textrm{c}}<15\, \si{\kelvin}$ (class 0) and $T_{\textrm{c}}\geq15\, \si{\kelvin}$ (class 1) among the 85\% of the materials - namely, the training set - thus obtaining the pair of 2 dimensional binnings in Fig.~\ref{fig:qeg-metallic}a. 
For each of those binnings, we have computed the five needed constraints, namely the means of those two features and their three variance terms (see Methods).
We have thus constructed a surface of maximum Shannon entropy for each of the two classes by means of the QEG algorithm, as depicted in Fig.~\ref{fig:qeg-metallic}b.
Finally, we have computed the probability distribution by subtracting the QEG solution for class 0 from the QEG solution for class 1 - both multiplied by the cardinality of the corresponding class in the training set - and up-shifting the result by the minimum, in such a way to have probabilities $\geq0$. 
The latter distribution represents our 2 dimensional QEG probabilistic classifier. 
Moreover, having in mind the idea of Random Forests, which employ \textit{bagging} (creation of more decision trees and aggregation of the results by taking the mean) \cite{hastie2009random}, we have produced 100 QEG 2D distributions per class, each based on a different random subset of training set. 
We have thus taken a mean distribution per class; Fig.~\ref{fig:qeg-metallic}c
shows that the bagged results are in accordance with the \emph{non-bagged} case of Fig.~\ref{fig:qeg-metallic}b.

\begin{figure}
    \centering
    \includegraphics[width = 0.45\textwidth]{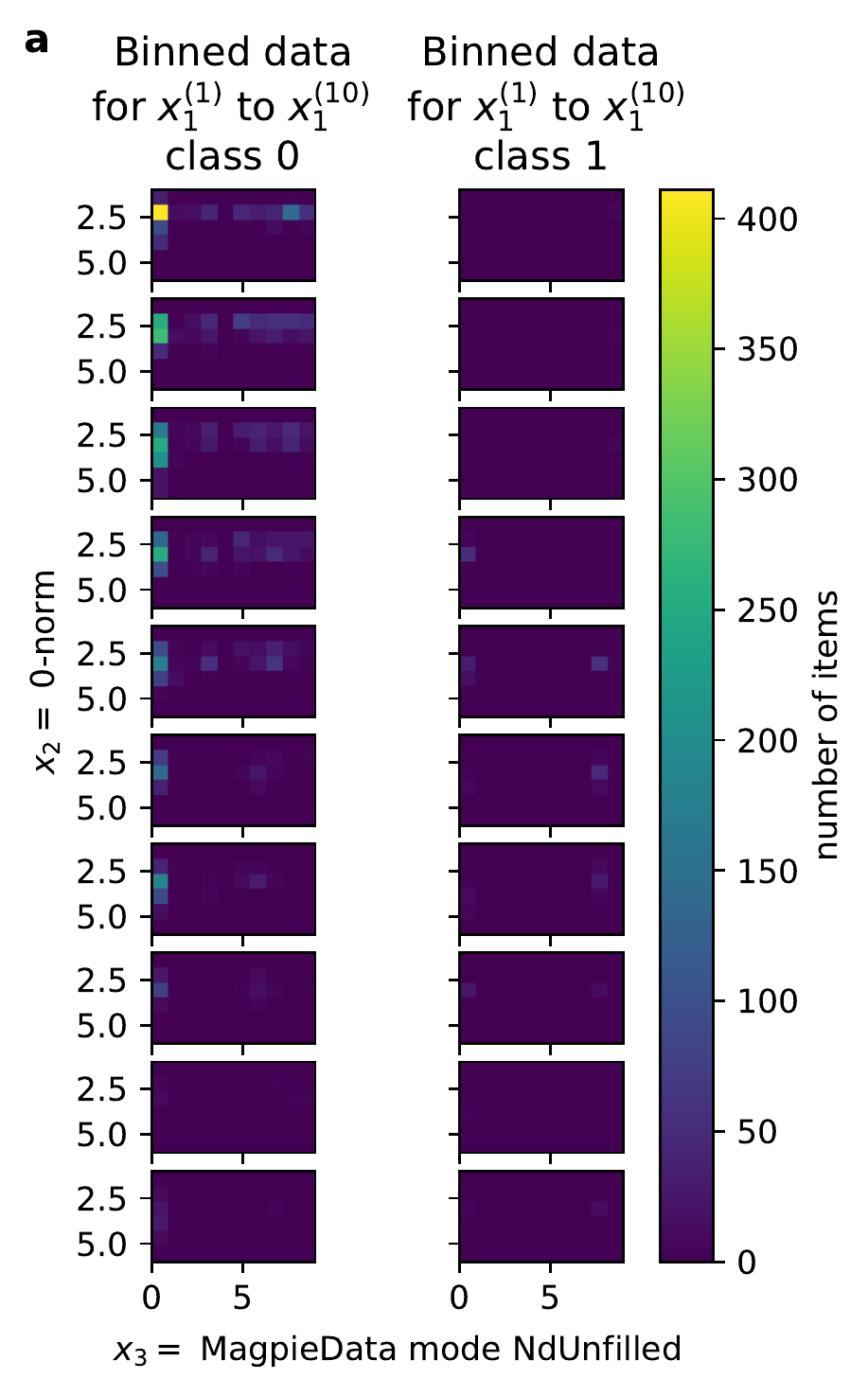}
    \includegraphics[width = 0.47\textwidth]{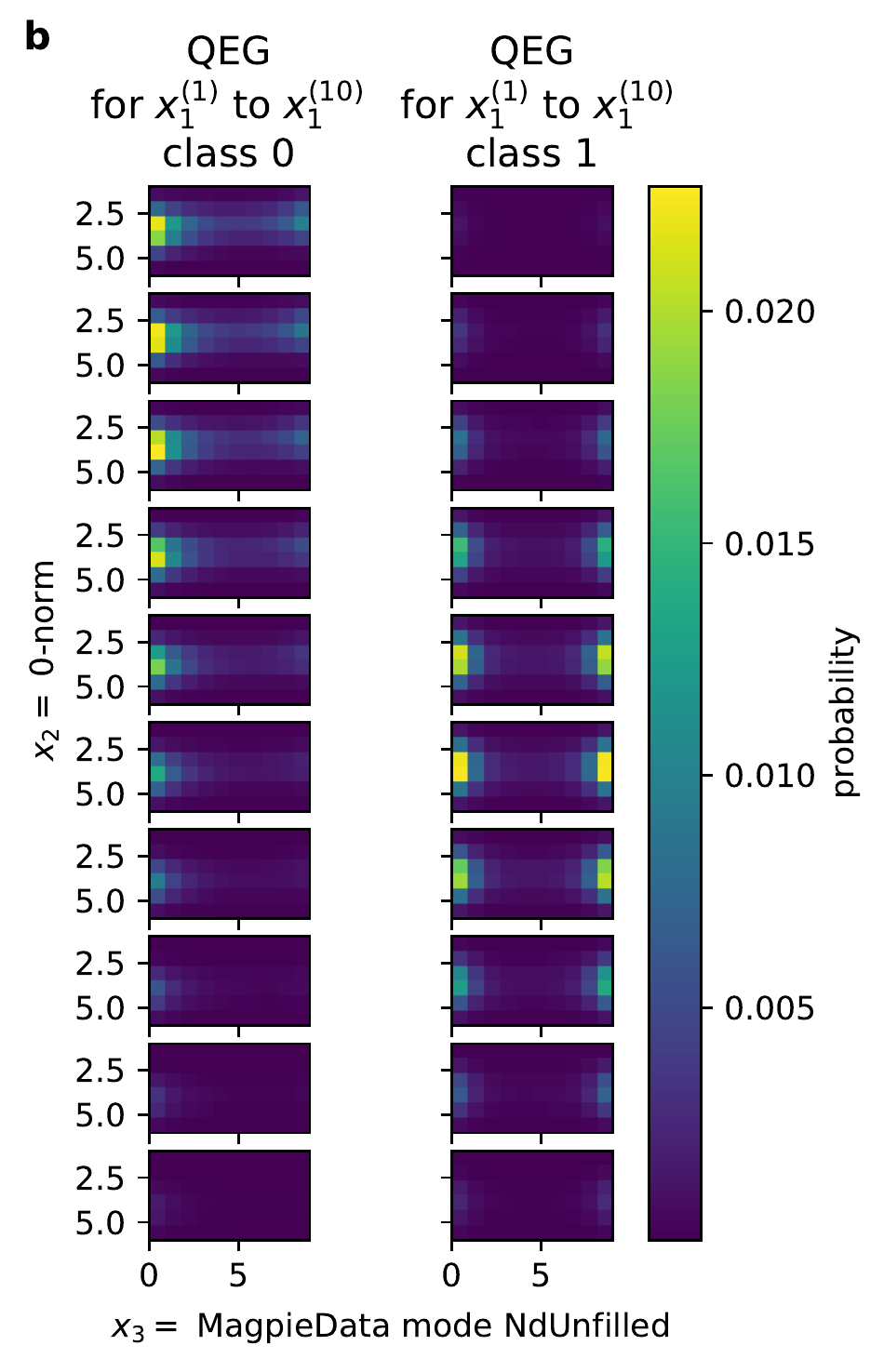}
    \caption{Probabilistic classifier. \textbf{a} 3-dimensional binning, with 10 bins for the first variable, 5 for the second, 10 for the third, of the two most relevant features $x_1, x_2, x_3$ according to the SHAP ranking for superconductors showing $T_{\rm{c}}<15\, \si{\kelvin}$ and $T_{\rm{c}}\geq15\, \si{\kelvin}$ respectively; \textbf{b} QEG solution of corresponding maximum Shannon entropy probability distribution.}
    \label{fig:qeg-metallic-3d}
\end{figure}

We have repeated the same procedure taking into account the first three features according to the SHAP-based ranking above - namely, the range of the melting temperature, the number of different chemical species, the mode of $d$ unfilled orbitals.
A 3-dimensional binning of dimensions $10\times6\times10$ can be represented as an ensemble of ten 2-dimensional binnings, each of dimensions $6\times10$.
Fig.~\ref{fig:qeg-metallic-3d}a shows such discretization of the data, where ten matrices of axes $x_2, x_3$ act for the ten bins of feature $x_1$, from bin $x_1^{(1)}$ to $x_1^{(10)}$ for each of the two classes. 
Fig.~\ref{fig:qeg-metallic-3d}b shows the corresponding the QEG solutions for classes 0 and 1 separately.

\subsection{Other standard binary classifiers}
Furthermore, we have trained and validated two Extra Trees Classifier (ETC) models with default hyperparameters over the same training set accounting for the 85\% of materials, including only the first two and the first three features by the aforementioned SHAP ranking respectively.
Moreover, we have used the entire dataset - with all the features - to train and validate two further ETC-based pipelines, both with pre-processing and hyperparameter tuning in stratified 5-fold cross validation (see Supplementary Note 4 for details). 
Specifically, since the cardinalities of two classes are unbalanced, we employed the Synthetic Minority Over-Sampling TEchnique (SMOTE) algorithm in one of the two pipelines, which, through interpolation, produces samples in the underrepresented class \cite{chawla2002smote}. 

In particular, the Scikit-learn Python package \cite{scikit-learn} offers the possibility of predicting not only the class, but also the class probabilities; the predicted class is automatically chosen to be the one accounting for the highest probability. 
Hence, by considering only the probabilities of the class 1, i.e., the material is predicted to be superconductive, we have moved the discriminating threshold from 0 (all the materials are predicted in class 1) to 1 (all the materials are predicted in class 0). 
For each threshold, a different confusion matrix, with different number of true positives ($\rm{TP}$), false negatives ($\rm{FN}$), false positives ($\rm{FP}$), true negatives ($\rm{TN}$), is constructed. 
For each confusion matrix, the true positive rate ($\rm{TPR}$) and the false positive rate ($\rm{FPR}$) are computed, where $\rm{TPR} = \rm{TP}/(\rm{TP}+\rm{FN})$ and $\rm{FPR} = \rm{FP}/(\rm{FP}+\rm{TN})$. The same procedure is repeated for the QEG based probabilistic classifiers, where the order of magnitude of the thresholds is lower, since the probability does not sum up to 1 over two classes but over 60 (QEG 2D) or 600 bins (QEG 3D).

 In Supplementary Note 8 we report a comprehensive comparison of the Receiver Operating Characteristic (ROC) curves for the all the classifiers.

The performence of a classifier can be measured by means of the Area Under Curve (AUC) of the ROC: the larger the AUC, the better the classifier.
%
%
Furthermore, given a ROC curve, its best discriminating threshold $\xi$ - above which a sample is classified as 1 and below which is classified as 0 - can be identified by means of the Youden's statistics, maximizing the quantity $J=\rm{TPR}-\rm{FPR}$ \cite{youden1950index}. 
Another metric for choosing the best threshold is the maximization of the $F_1$ score, by definition $F_1=2\rm{TP}/(2\rm{TP} + \rm{FP} + \rm{FN})$ \cite{chinchor1992proceedings, van1977theoretical, taha2015metrics}. 

Performances computed over the same testing set of 1084 materials are shown in Table \ref{tab:performances}.  
The comparison encompasses QEGs with two and three features (QEG 2D and QEG 3D respectively), ETCs with the top two and three features of the SHAP ranking (ETC 2D-high and ETC 3D-high respectively), ETCs with the two (33-rd, 34-th) and three (33-rd, 34-th, 35-th) features of the SHAP ranking (ETC 2D-middle and ETC 3D-middle respectively), with the least two and three features of the SHAP ranking (ETC 2D-low and ETC 3D-low respectively), ETC with all the database (ETC-vanilla), ETC with the additional SMOTE algorithm (ETC-SMOTE), ETC with all the database and all the 81 features (ETC-vanilla-81), ETC with the additional SMOTE algorithm and all the 81 features (ETC-SMOTE-81), Gaussian Naive Bayesian classifier (Naive 2D, see Supplementary Note 10 and ref.~\cite{zhang2004optimality} for details), together with a \emph{No skill} classifier, in which $\rm{TPR}$ and $\rm{FPR}$ are always equal.
ETC models always outperform QEG-based classifiers both in terms of $J_{\rm{max}}$ and in terms of $F_{\rm{1,max}}$; in particular, the ETC-vanilla and ETC-SMOTE turn out to be the best classifiers in terms of $F_{\rm{1,max}}$ and $J_{\rm{max}}$ respectively. 

\begin{table*}
\centering
\caption{Performances of the trained classifiers.}
\label{tab:performances}
\begin{tabular}{lrrrrr}
\toprule
& AUC & $\xi_{J, \rm{max}}$ & $J_{\rm{max}}$ & $\xi_{F_1,\rm{max}}$ &  $F_{1, \textrm{max}}$\\
\midrule
No skill & 0.50 & - & - & - & - \\
QEG 2D & 0.71 & 0.004 & 0.40 & 0.017 & 0.23 \\
QEG 2D bagged & 0.71 & 0.004 & 0.40 & 0.017 & 0.23 \\
QEG 3D & 0.60 & 0.002 & 0.29 & 0.002 & 0.22 \\
ETC 2D-high & 0.96 & 0.120 & 0.90 & 0.313 & 0.73 \\
ETC 3D-high & 0.96 & 0.125 & 0.89 & 0.333 & 0.73 \\
ETC 2D-middle & 0.86 & 0.028 & 0.63 & 0.317 & 0.38 \\
ETC 3D-middle & 0.82 & 0.167 & 0.62 & 0.167 & 0.52 \\
ETC 2D-low & 0.54 & 0.072 & 0.08 & 0.072 & 0.14 \\
ETC 3D-low & 0.54 & 0.073 & 0.08 & 0.073 & 0.14 \\
ETC-vanilla & 0.99 & 0.110 & 0.91 & 0.560 & 0.85\\
ETC-SMOTE & 0.98 & 0.216 & 0.92 & 0.780 & 0.83\\
ETC-vanilla-81 & 0.98 & 0.040 & 0.91 & 0.630 & 0.84\\
ETC-SMOTE-81 & 0.99 & 0.140 & 0.91 & 0.732 & 0.84\\
Naive 2D & 0.85 & 0.071 & 0.73 & 0.134 & 0.37\\

\bottomrule
\end{tabular}
\end{table*}

We have thus predicted the probability of classes 0 and 1 with ETC-vanilla and ETC-SMOTE for all the $\sim 40,000$ materials in Materials Project without \ch{Ni}, \ch{Fe}, \ch{Cu}, \ch{O} and not in the SuperCon database. Those predictions are on our GitHub repository (see Code availability). 

\subsection{Optimal reduction of the composition-based material descriptors}
Although the above SHAP analysis can be conveniently adopted while ranking and reducing the number of material descriptors for both regressors and classifiers, the following aspects have to be stressed.
On one hand, as visible on the right-hand side of Fig.~\ref{fig:model}, for achieving a sufficiently high (i.e. in the order of 70\% or higher) cumulative importance over 30 features are needed.
On the other hand, the larger the number of feature the higher the over-fitting possibility.
Therefore, in this work, we attempted the following possible reduction of the material descriptors.
Given the original set of $n$ features $(x_1,...,x_n)$, let $(\tilde{x}_1,...,\tilde{x}_n)$ be the corresponding dimensionless quantities:
\begin{equation}\label{normalizedfeature}
    \tilde{x}_i=\frac{x_i-x_{i,\rm{min}}}{x_{i,\rm{max}}-x_{i,\rm{min}}}+1
\end{equation}
where $x_{i,\rm{min}}$ and $x_{i,\rm{max}}$ represent the minimum and maximum observed values for the $i-$th feature over the training set, respectively.
All dimensionless quantities are thus normalized by construction to a value range within the interval $[1-2]$ to avoid singularities in the expressions below.

We define the following new set of $m\ll n$ mixed features $(y_1,...,y_m)$, as follows:
\begin{equation}\label{mixedpower}
    y_j=\prod_{i=1}^n \tilde{x}_i^{\alpha_{ij}}
\end{equation}
where $\{\alpha_{ij}\}$ represents an $n \times m$ matrix optimally estimated as reported below.
Alternatively, the new set of $m$ reduced mixed features can also be defined by the following linear transformation:
\begin{equation}\label{mixedlinear}
    y_j=\sum_{i=1}^n \alpha_{ij} \tilde{x}_i
\end{equation}
Finally, the new variables $y_j$ can be conveniently normalized within the interval $[0-1]$ as follows:
\begin{equation}
    \tilde{y}_j=\frac{y_j-y_{j,\rm{min}}}{y_{j,\rm{max}}-y_{j,\rm{min}}}
    \end{equation}
With the basic idea of Bayesian classification in mind, we define the following multi-objective optimization criterion.
The matrix $\{\alpha_{ij}\}$ in Eq.~\ref{mixedpower} and/or Eq.~\ref{mixedlinear} lies on the Pareto front while concurrently attempting: i) maximization of a properly chosen distance between the two classes; ii) minimization of a norm of the covariance matrix of the first class distribution; iii) minimization of a norm of the covariance matrix of the second class distribution.

In this study, we use genetic algorithms for optimization. 
Moreover, for the evaluation of the distance between the two classes, a number of approaches have been tested including:
\begin{itemize}
\item Data in the two classes are equally binned and histograms used to evaluated the Bhattacharyya distance \cite{bhattacharyya1943measure, bhattacharyya1946measure} between the two classes to be maximized during the above multi-objective optimization;
\item Data in the two classes are equally binned and histograms used to evaluate the Earth mover distance \cite{villani2009optimal} between the two classes to be maximized during the above multi-objective optimization;
\item The average number of neighbors within a fixed radius of non superconducting materials to each sample of the superconducting material class in the reduced space to be minimized during the above multi-objective optimization
\end{itemize}
Finally, for the remaining two objective functions, while for one-dimensional cases a numerical estimate of the standard deviation of the binned data in the two classes is computed, in the two (or higher) dimensional cases the determinant of the covariance matrix can be adopted. 
More details about Pareto front calculations can be found in Supplementary Note 9.

\subsubsection{Application to one- and two-dimensional cases}
%
%
\begin{figure}
    \centering
    \includegraphics[width = 0.9\textwidth]{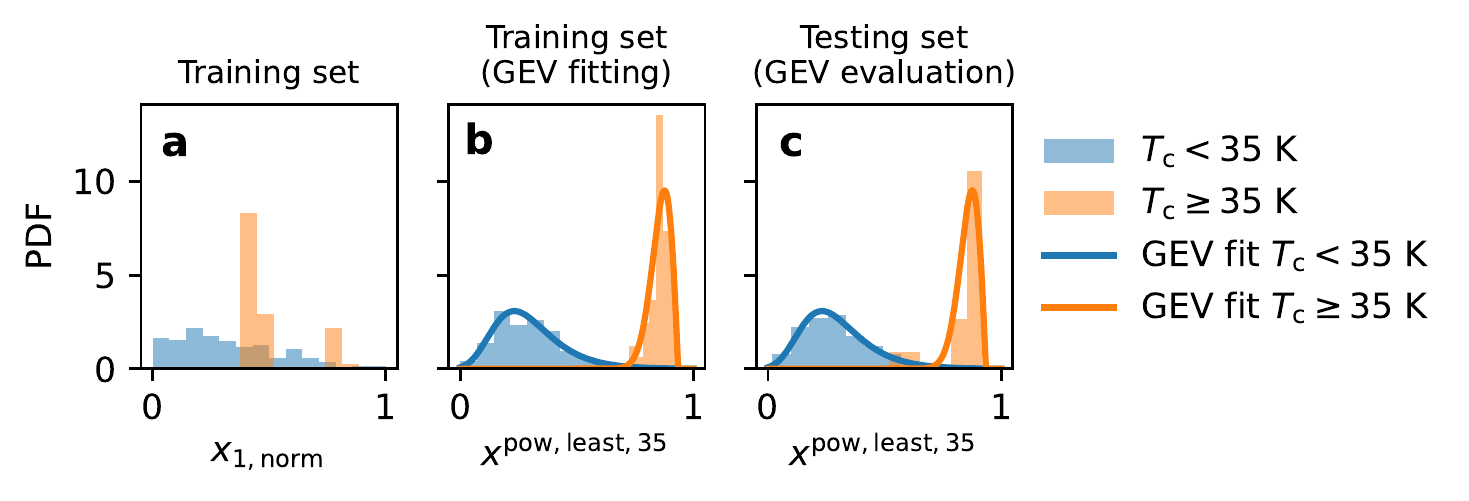}
    \caption{One-dimensional example. \textbf{a}: PDFs over binned data of the training set for the two classes ($T_{\rm{c}}<35\, \si{\kelvin}$ and $T_{\rm{c}} \geq 35\, \si{\kelvin}$) reported against the normalized first most important feature according to the SHAP ranking. \textbf{b}: PDFs over binned data of the training set for the two classes reported against the mixed feature $x^{\rm{pow, least, 35}}$, constructed according to Eq.~\ref{mixedpower} and choosing the point of the Pareto front with the least overlapping of the two classes according to the Battacharyya distance, together with a GEV analytical fitting of those two binnings (see text for details). \textbf{c}: PDFs over binned data of the testing set for the two classes reported against the same mixed feature $x^{\rm{pow, least, 35}}$ together with the same GEV fittings of the \textbf{b} subfigure.
}
    \label{1Dmixed}
\end{figure}
    

%
%
\begin{figure}
    \centering
    \includegraphics[width = 0.7\textwidth]{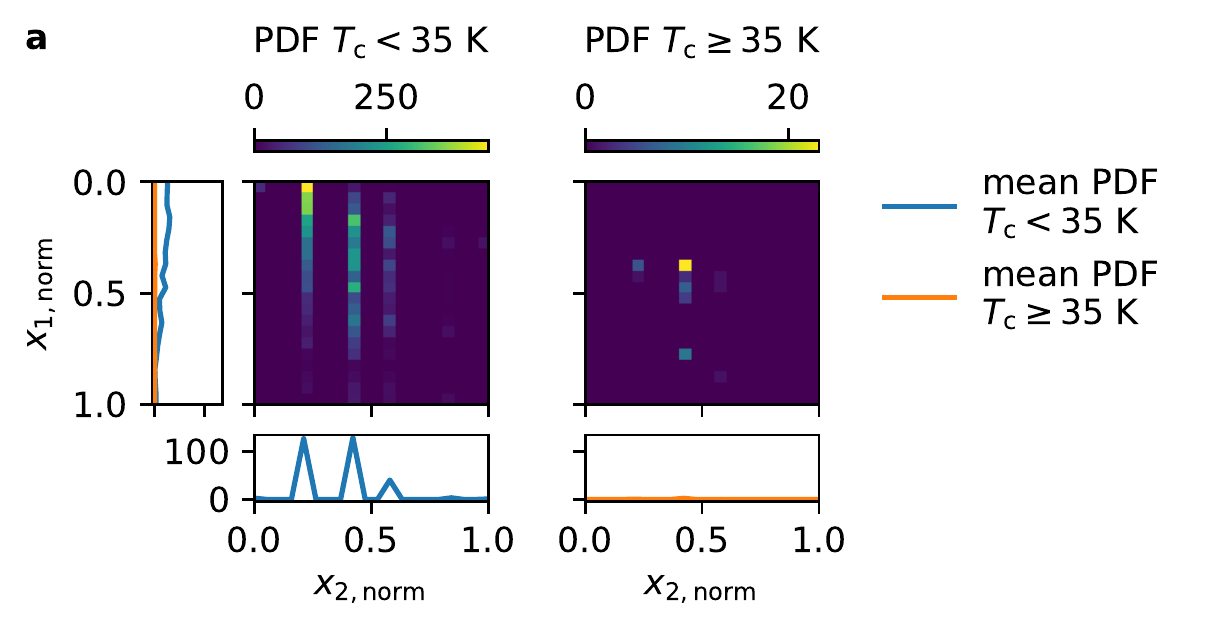}
    \includegraphics[width = 0.7\textwidth]{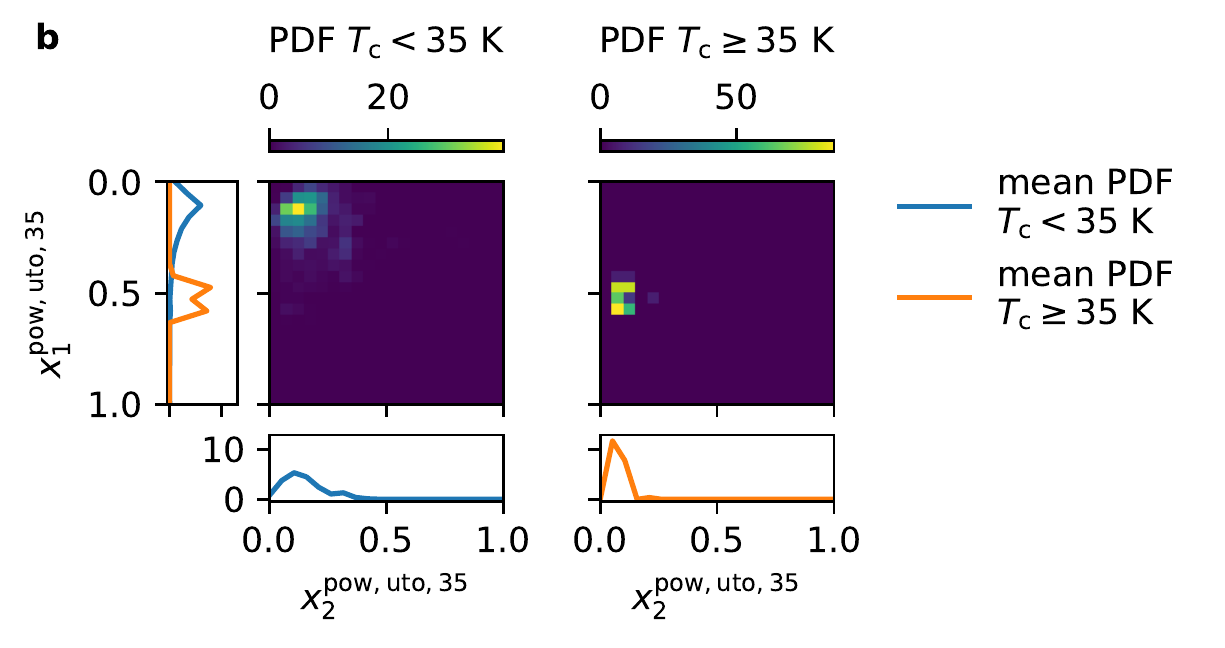}
    \includegraphics[width = 0.7\textwidth]{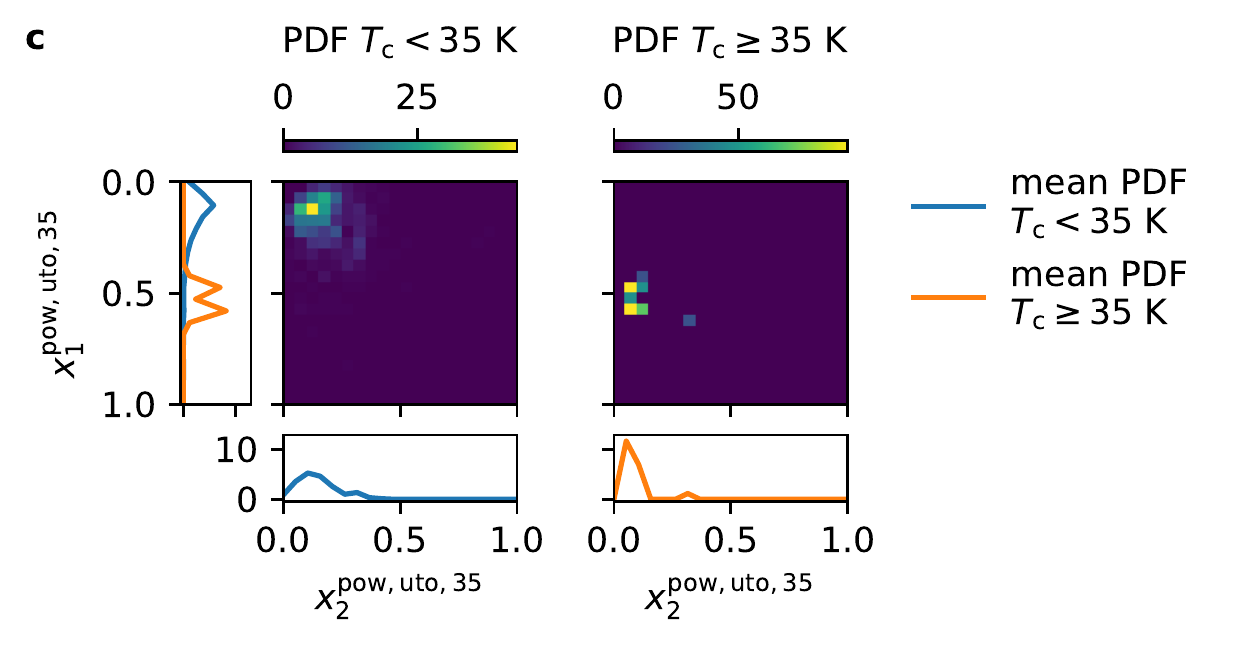}
    
    \caption{Two dimensional example. \textbf{a}: PDFs over binned data of the training set for the two classes ($T_{\rm{c}}<35\, \si{\kelvin}$ and $T_{\rm{c}} \geq 35\, \si{\kelvin}$) reported against the normalized first most important feature according to the SHAP ranking. \textbf{b}: PDFs over binned data of the training set for the two classes reported against the two mixed features $x_1^{\rm{pow, uto, 35}}$ and $x_2^{\rm{pow, uto, 35}}$, constructed according to Eq.~\ref{mixedpower} from mixing the 52 most important features according the the SHAP ranking and choosing the Utopia point of the Pareto front. \textbf{c}: PDFs over binned data of the testing set for the two classes reported against the same mixed features $x_1^{\rm{pow, uto, 35}}$ and $x_2^{\rm{pow, uto, 35}}$.}
    \label{2Dmixed}
\end{figure}

As an example, Fig.~\ref{1Dmixed} shows the Probability Density Functions (PDFs) of the two material classes $T_{\rm{c}}<35\, \si{\kelvin}$ and $T_{\rm{c}} \geq 35 \, \si{\kelvin}$. 
Specifically, Fig.~\ref{1Dmixed}a reports the PDF binning of the training set data over the two classes, against the normalized most important feature according to the SHAP ranking. 
Fig.~\ref{1Dmixed}b shows the same PDFs against the mixed feature $x^{\rm{pow, least, 35}}$, constructed according to Eq.~\ref{mixedpower} by power combination of the 30 most important features of the SHAP ranking and choosing the point of the Pareto front with the least distributions overlap according to the Bhattacharyya distance.

%
%
%
Interestingly, when plotted against the new mixed feature, the two classes  appear well separated, whereas it is worth observing that the same two classes show a higher degree of overlapping when reported against the first SHAP feature.
As a result, it appears particularly convenient to attempt an analytical bet-fitting of the two functions reported Fig.~\ref{1Dmixed}b, approximated by a Generalized Extreme Value (GEV) distribution, whose density has equation
\begin{equation}
g\left(x^{\rm{pow, least, 35}}\right) = \frac{1}{\sigma}\left(1+\zeta\frac{x^{\rm{pow, least, 35}}-\gamma}{\sigma}\right)^{-\frac{\zeta+1}{\zeta}}\exp\left(-\left(1+\zeta\frac{x^{\rm{pow, least, 35}}-\gamma}{\sigma}\right)^{-1/\zeta}\right).
\end{equation}
In this specific case, we found that the GEV distribution for materials with $T_{\rm{c}}<35\, \si{\kelvin}$ has factors $\gamma=0.228$, 
$\sigma=0.119$, $\zeta=-0.033$. 
Analogously we found that the GEV distribution for materials with $T_{\rm{c}} \geq 35\, \si{\kelvin}$ has factors $\gamma=0.847$, 
$\sigma=0.046$, $\zeta=-0.539$. We performed such fittings by means of the SciPy Python package \cite{2020SciPy-NMeth}. 
Fig.~\ref{1Dmixed}c shows the PDFs over the binned data of the testing set reported against the same mixed feature, together with the GEV fittings computed on the training set. 
It is worth noticing that the classes are still well separated, with a good agreement between the GEV distributions and the testing set densities.
The number of bins has been chosen separately for the two classes, according to the Sturges rule \cite{sturges1926choice}.

Furthermore, Fig.~\ref{2Dmixed} shows the PDFs of the same two material classes ($T_{\rm{c}}<35\, \si{\kelvin}$ and $T_{\rm{c}} \geq 35 \, \si{\kelvin}$) in a two dimensional case. 
Specifically, Fig.~\ref{2Dmixed}a reports the PDF two dimensional binning of the training set data over the two classes, against the normalized two most important features according to the SHAP ranking. 
Fig.~\ref{2Dmixed}b shows the same PDFs against the mixed features $x_1^{\rm{pow, uto, 35}}, x_2^{\rm{pow, uto, 35}}$, constructed according to Eq.~\ref{mixedpower} by power combination of the 52 most important features of the SHAP ranking and choosing the Utopia point of the Pareto front. 
As in the one dimensional case, the two classes, when plotted against the new mixed features, appear well separated. Fig.~\ref{2Dmixed}c shows the PDFs over the binned data of the testing set reported against the same mixed features; the two classes are still well separated. 
Each plot of Fig.~\ref{2Dmixed} accounts for 400 two dimensional bins, on a grid $20\times20$.
%
Moreover, Supplementary Note 11 shows a sharp improvement of a Naive Gaussian Bayesian classifier trained with the mixed features $x_1^{\rm{pow, uto, 35}}$, $x_2^{\rm{pow, uto, 35}}$ with respect to an analogous model trained with the two most relevant features according to the SHAP ranking.

All the relevant data of the Pareto fronts used for constructing those mixed features, together with the coefficients $\alpha_{ij}$ of each case, are publicly available on our GitHub repository (see Code availability).


\subsection{Possible generalizations}
\begin{figure}
    \centering
    \includegraphics[width = 0.7\textwidth]{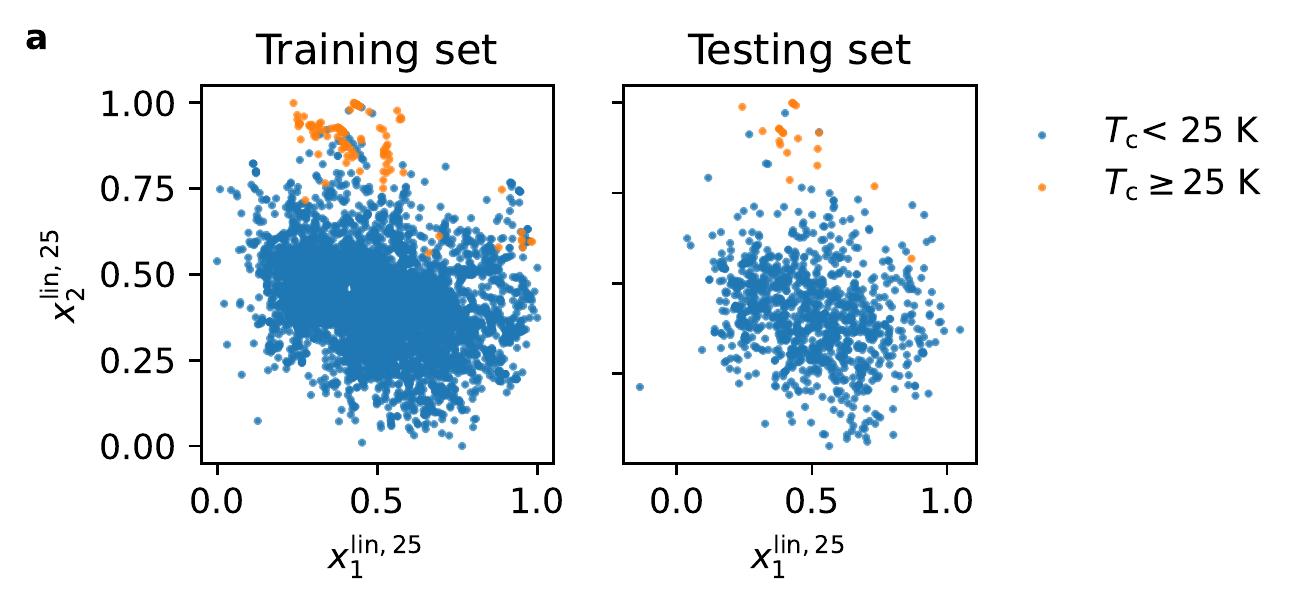}
    \includegraphics[width = 0.7\textwidth]{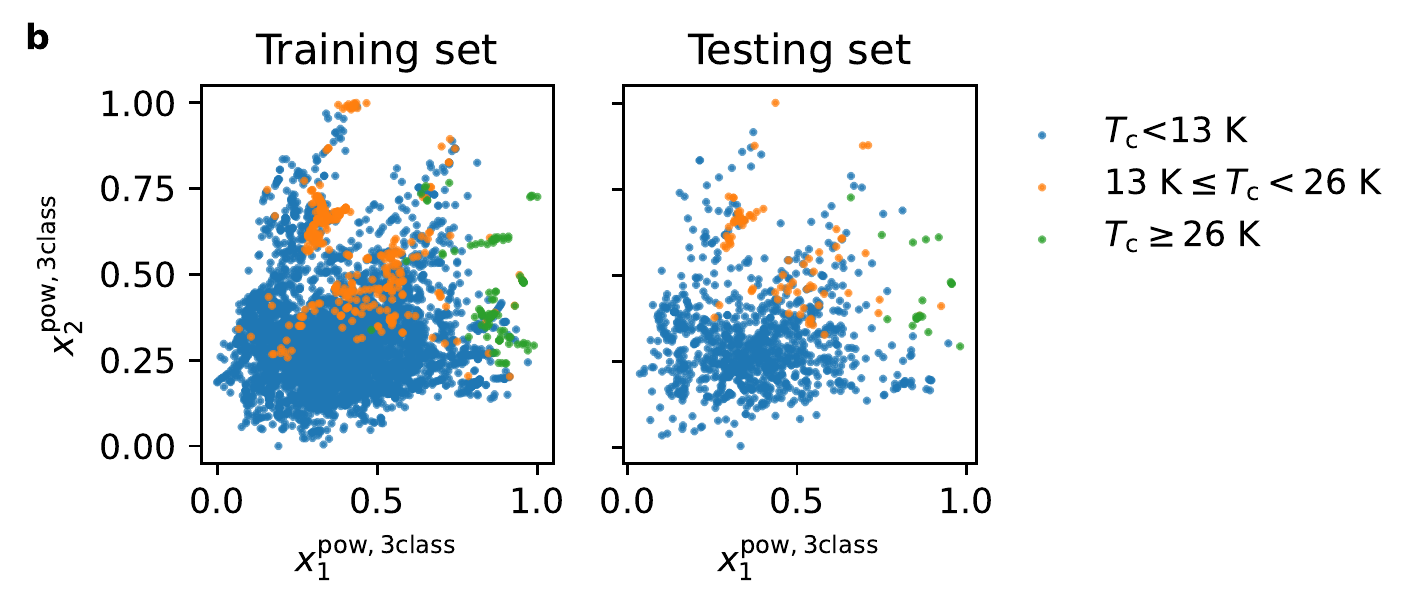}
    \caption{Projections of training and testing sets into the reduced feature space with colors indicating the critical temperature classes. \textbf{a} Projection over the two mixed features $x_1^{\rm{lin, 25}}$,  $x_2^{\rm{lin, 25}}$, constructed according to Eq.~\ref{mixedlinear} and obtained by single objective optimization, where the Bhattacharyya distance between the two classes $T_{\rm{c}}<25\, \si{\kelvin}$ and $T_{\rm{c}}\geq25\, \si{\kelvin}$ has been maximized. \textbf{b} Projection over the two mixed features $x_1^{\rm{pow, 3class}}$, $x_2^{\rm{pow, 3class}}$, constructed according to Eq.~\ref{mixedpower} and obtained by multi-objective optimization where the Bhattacharyya pairwise distances between the three classes $T_{\rm{c}}<13\, \si{\kelvin}$, $13\, \si{\kelvin}\leq T_{\rm{c}}<26\, \si{\kelvin}$, $T_{\rm{c}}\geq26\, \si{\kelvin}$ have been concurrently maximized.}
    \label{singleopt}
\end{figure}

We are conscious that the mixed features found in this work might be still sub-optimal, as we do not have here the ambition of comprehensively exploring all possible cases.
Clearly, several generalizations and variations can be studied while performing the material descriptor reduction as discussed above.
Obvious generalizations might adopt different functions for reducing variables as compared to equations \ref{mixedpower} and \ref{mixedlinear}, as well as different distance functions between the classes.
Alternatively, other strategies for constructing optimal mixed features might also focus on distances only between classes thus neglecting minimization of variance terms, with the primary aim being the best separation between classes.
In this respect, we report the following examples:
\begin{itemize}
    \item The training dataset is split in two classes (i.e. materials with a critical temperature above or below a certain threshold value) and a single objective optimization is performed only aiming at maximizing the distance between two classes (see Fig.~\ref{singleopt}a);
    \item The training dataset is split in multiple classes (i.e. $>2$) and a multi-objective optimization is performed aiming at concurrently maximing the pairwise distances between the classes (see Fig.~\ref{singleopt}b).
\end{itemize}
For further details, please refer to Supplementary Note 9.

%
%

\subsection{Entropy-, tree-, and Bayes-based binary classifiers on the new mixed features}
We thus report the results of the QEG-based probabilistic classifiers and ETCs by employing the new mixed features, always constructed by aggregating the top 30 features of the SHAP ranking, in both the cases of power (Eq.~\ref{mixedpower}) and linear (Eq.~\ref{mixedlinear}) transformations.
In these examples, for the purposes of optimization and class separation, we consider only the Utopia point of the Pareto front and the Bhattacharyya distance respectively. 

\begin{figure}
    \centering
    \includegraphics[width = 0.48\textwidth]{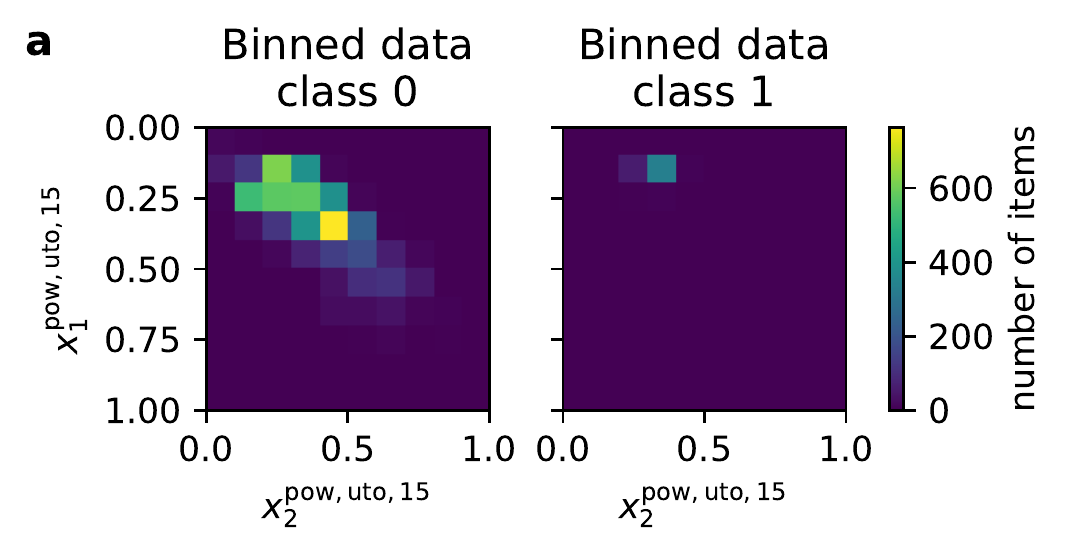}
    \includegraphics[width = 0.48\textwidth]{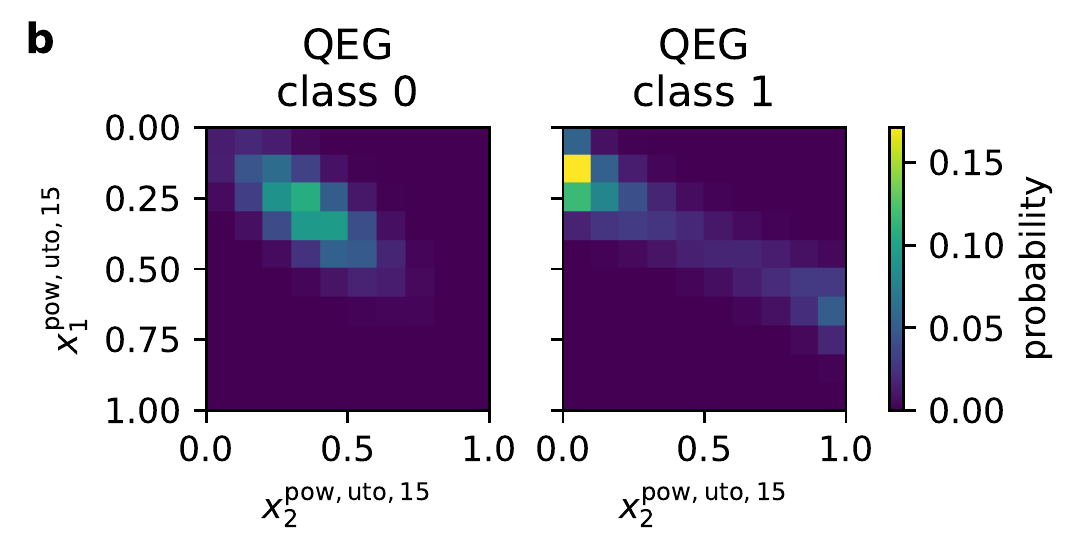}
    \caption{Probabilistic classifier. \textbf{a} 2-dimensional binning, with 10 bins for each variable, of the two mixed features $x^{\rm{pow, uto, 15}}_1$, $x^{\rm{pow, uto, 15}}_2$ constructed according to Eq.~\ref{mixedpower}, by selecting the utopia point of the Pareto front, for superconductors showing $T_{\rm{c}}<15\, \si{\kelvin}$, and $T_{\rm{c}}\geq15\, \si{\kelvin}$ respectively; \textbf{b} QEG solution of corresponding maximum Shannon entropy probability distribution.}
    \label{fig:qeg-metallic-2d-pow}
\end{figure}

\begin{figure}
    \centering
    \includegraphics[width = 0.48\textwidth]{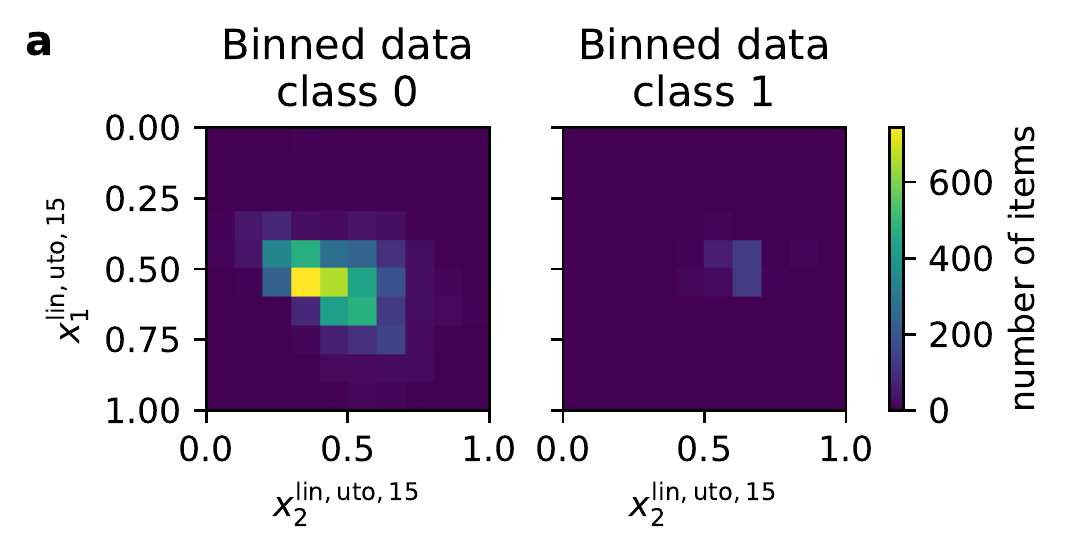}
    \includegraphics[width = 0.48\textwidth]{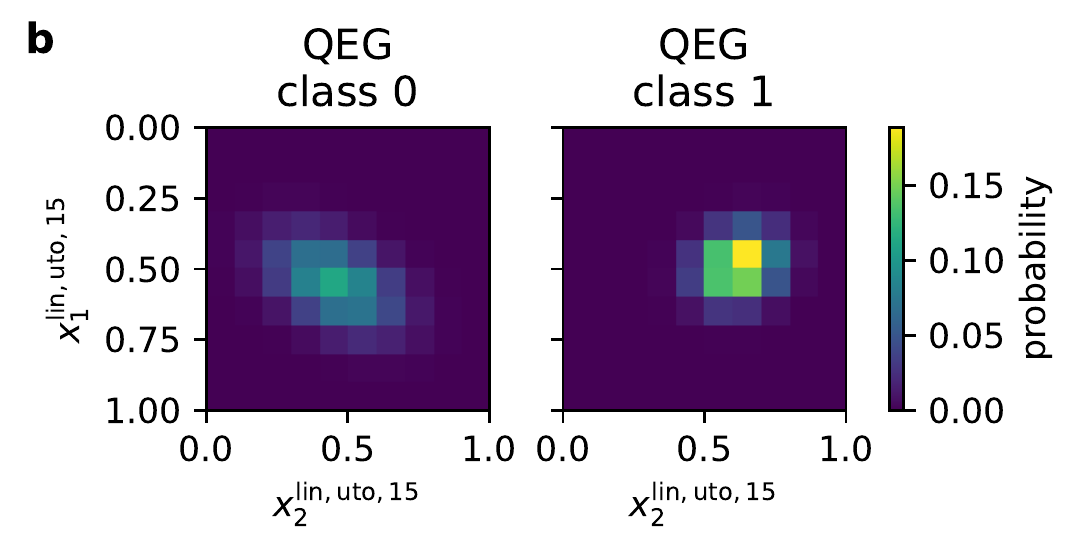}
    \caption{Probabilistic classifier. \textbf{a} 2-dimensional binning, with 10 bins for each variable, of the two mixed features $x^{\rm{lin, uto, 15}}_1, x^{\rm{lin, uto, 15}}_2$ constructed according to Eq.~\ref{mixedlinear},by selecting the utopia point of the Pareto front, for superconductors showing $T_{\rm{c}}<15\, \si{\kelvin}$ and $T_{\rm{c}}\geq15\, \si{\kelvin}$ respectively; \textbf{b} QEG solution of corresponding maximum Shannon entropy probability distribution.}
    \label{fig:qeg-metallic-2d-lin}
\end{figure}

\begin{table*}
\centering
\caption{Performances of the classifiers trained with mixed features.}
\label{tab:performances_mixed}
\begin{tabular}{lrrrrr}
\toprule
& {AUC} & $\xi_{J, \rm{max}}$ & $J_{\rm{max}}$ & $\xi_{F_1,\rm{max}}$ &  $F_{1, \textrm{max}}$\\
\midrule
No skill & 0.50 & - & - & - & - \\
QEG 2D-mixed pow & 0.69 & 0.007 & 0.52 & 0.007 & 0.29\\
QEG 2D-mixed lin & 0.79 & 0.008 & 0.61 & 0.011 & 0.37 \\
ETC 2D-mixed pow & 0.95 & 0.104 & 0.82 & 0.480 & 0.72 \\
ETC 2D-mixed lin & 0.93 & 0.092 & 0.77 & 0.574 & 0.70 \\
Naive 2D-mixed pow & 0.94 & 0.506 & 0.76 & 0.504 & 0.67 \\
Naive 2D-mixed lin & 0.90 & 0.502 & 0.78 & 0.505 & 0.50 \\

\bottomrule
\end{tabular}
\end{table*}

Specifically, Figs.~\ref{fig:qeg-metallic-2d-pow}a and \ref{fig:qeg-metallic-2d-pow}b show the binnings of the two classes ($T_{\rm{c}}<15\, \si{\kelvin}$ and $T_{\rm{c}}\geq15\, \si{\kelvin}$) against the two power mixed features $x_1^{\rm{pow, uto, 15}}$, $x_2^{\rm{pow, uto, 15}}$ and the corresponding QEG solution respectively. 
Analogously, Figs.~\ref{fig:qeg-metallic-2d-lin}a and \ref{fig:qeg-metallic-2d-lin}b show the binnings of the two classes ($T_{\rm{c}}<15\, \si{\kelvin}$ and $T_{\rm{c}}\geq15\, \si{\kelvin}$) against the two linear mixed features $x_1^{\rm{lin, uto, 15}}$, $x_2^{\rm{lin, uto, 15}}$ and the corresponding QEG solution respectively.
Table~\ref{tab:performances_mixed} reports the performances of such classifiers, ending up with a consistent improvement of both the $J_{\rm{max}}$ and the $F_{1, \rm{max}}$ score with respect to the case of the QEG 2D trained with the top SHAP descriptors (see QEG 2D in Table~\ref{tab:performances}); specifically, the linear transformation improves also the AUC. 
We have used the same mixed features to train and validate two ETCs, ending up with similar metrics - AUC, $J_{\rm{max}}$ and $F_{1, \rm{max}}$ - score with respect to the case ETC 2D-high, trained with the two most relevant features according to the SHAP ranking. 
We have finally employed the same features to re-train also the Gaussian Bayesian Classifier, getting an improvement for all the metrics (namely, AUC, $J_{\rm{max}}$ and $F_{\rm{1, max}}$) with respect to the same classifier trained with the top two features according to the SHAP ranking, both for the power transformation and for the linear transformation.

The corresponding ROC curves are reported in Supplementary Note 8. 


\section{Conclusions}
%

Here we have developed several ML tools for studying the critical temperature of superconductors. From the SuperCon database, we have considered only the inorganic compounds without \ch{Fe}, \ch{Ni}, \ch{Cu}, \ch{O}, thus excluding oxides that belong to low temperature classic superconductors. 
By means of Matminer and on the basis of the SuperCon database, we have generated 145 composition-based features for each compound. 
We have trained and validated a tree-based regression model for the prediction of the $T_{\rm{c}}$, allowing us to identify the most relevant descriptors by means of the Tree SHAP routine. 
Then, we have produced several different classifiers, based on different sets of features and considering materials with $T_{\rm{c}}\geq15\, \si{\kelvin}$ in class 1 and materials with $T_{\rm{c}}<15\, \si{\kelvin}$ in class 0.
%
In particular, with the idea of Bayesian classifiers in mind, we have tested a new Entropy-based classifier (here referred to as QEG), approximating the multidimensional binning of the data over the chosen set of descriptors with the surface of maximum Shannon Entropy. 
Other employed models include tree-based classifiers (namely ETCs) and Naive Bayesian classifiers. 
In particular, by comparing ETCs using only two or three of the original extracted features, we notice that the SHAP ranking - identified for regression - can be consistently used for classification. 
Since, ETCs with few features performed better than both QEGs and Naive Bayesian classifiers, we have trained two more comprehensive models - ETC-vanilla, ETC-SMOTE - both based on a number of features selected during the pre-processing routines of the respective ML pipelines. 
The latter uses also the SMOTE algorithm to sample, through interpolation, materials in the under-represented class of superconductors. 
Additionally, we have trained two further models - ETC-vanilla-81 and ETC-SMOTE-81, trained with the same ensemble of 81 features effectively used by the regression model ETR. We have employed the best-performing models, namely ETC-vanilla and ETC-SMOTE, to rank $\sim40,000$ compounds in MaterialsProject and not occurring in the Supercon, in terms of the probability of showing  $T_{\rm{c}}\geq15\, \si{K}$. For instance, ETC-vanilla predicts 41 of those formulae to show $T_{\rm{c}}\geq15\, \si{K}$ with probability not lower than $0.6$.
%
Furthermore, by means of multi-objective optimization, we have found optimized mixed features that proved particularly suitable for class separation. 
To this end, we have mixed by means of power or linear combination the top 30 features of the SHAP ranking. 
With such new features, the performances of both QEGs and Naive Bayesian classifiers improve, while the ETCs performances are in line with the corresponding models trained over the original features. 
Remarkably, in general there is no need to have access to the SHAP ranking for achieving such optimization, and, in principle, all the input features can be imported for mixing.

Additionally, we have produced further examples differing with the previous ones in terms of threshold $T_{\rm{c}}$ and/or optimization routines. 
Among those, we have found an optimal single feature to separate classes $T_{\rm{c}}<35\, \si{\kelvin}$ and $T_{\rm{c}}\geq35\, \si{\kelvin}$.
Interestingly, in this case we were able to give the equation of an analytical classifier fitted on the materials binned over new mixed feature. 
We have employed both the best QEG model - QEG 2D-mixed lin classifier (for $T_{\rm{c}}\geq15\, \si{\kelvin}$), and the analytical classifier (for $T_{\rm{c}}\geq35\, \si{\kelvin}$) to rank the same $\sim40,000$ materials of MaterialsProject not occurring in the SuperCon database. 
Such predictions are publicly available on our GitHub repository.

Another aim of this work was to test the possible invariance of the critical temperature with respect to binary groups of features in the form of $x_i^ax_j^b$. 
To this end, we have trained and validated a second regression model - i.e., a DNN - for the prediction of the $T_{\rm{c}}$, allowing us to compute the gradient of the critical temperature with respect to the input features, namely $\nabla T_{\rm{c}}(x_1, \dots, x_n)$.
Finally, we stress that the suggested methods in this papers, namely the search for invariant groups of regression models, the optimization of mixed composition based feature and the maximum entropy based classifiers are general and not restricted to the selected case study. 
As such we envision possible future applications to other energy materials such as thermal energy storage  \cite{aghemo2022comparison} and electrochemical energy storage \cite{wang2021deep} applications.

\section*{Data availability}
Processed datasatets and trained models of this study will be publicly available in Zenodo at (10.5281/zenodo.7725592) \cite{trezza_giovanni_2023_7725592}.

\section*{Code availability}
The codes used to obtain the results of this study will be publicly available in github at \url{https://github.com/giotre/superconductors}.

\section*{Acknowledgments}
The authors are grateful to Nicola Marzari (École Polytechnique Fédérale de Lausanne), Samuel Poncé (Université catholique de Louvain) and Marnik Bercx (École Polytechnique Fédérale de Lausanne) for the valuable discussions about the material selection in the case study reported in this work. E.C. acknowledges partial financial funding from the European Union's Horizon 2020 research and innovation programme under grant agreement No 957189.

\clearpage
\bibliographystyle{naturemag}
\bibliography{biblio}

\end{document}


\maketitle

\subsection*{Supplementary Note 1: Synthetic example for QEG}

We have generate a synthetic dataset with 5,000 samples consisting of pairs $(x_1, x_2)$ drawn from a bivariate Gaussian distribution, with mean vector and covariance matrix 
\begin{equation}
    \bm{\mu} = 
    \begin{pmatrix}
    -2\\1
    \end{pmatrix}, \hspace{.5cm}
    \bm{\Sigma} = \begin{pmatrix}
    1 & 2\\
    2 & 10\\
    \end{pmatrix}
\end{equation}
respectively. We have discretized such samples with 100 2-dimensional bins, as shown in Supplementary Fig.~\ref{fig:syntheticQEG}a; then, over this binning, we have constructed the QEG solution in Supplementary Fig.~\ref{fig:syntheticQEG}b.

\begin{figure}[H]
    \centering
    \includegraphics[width = 0.45\textwidth]{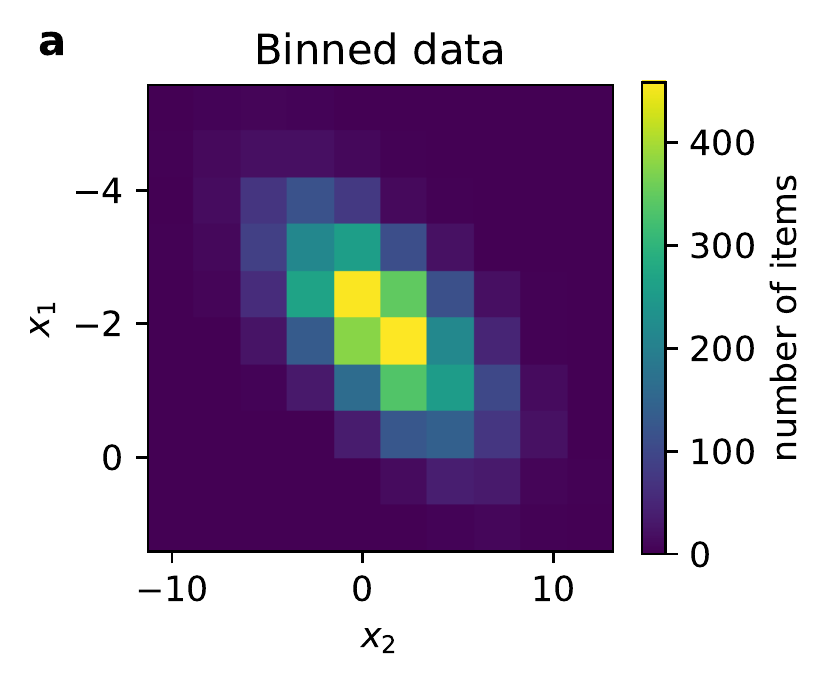}
    \includegraphics[width = 0.45\textwidth]{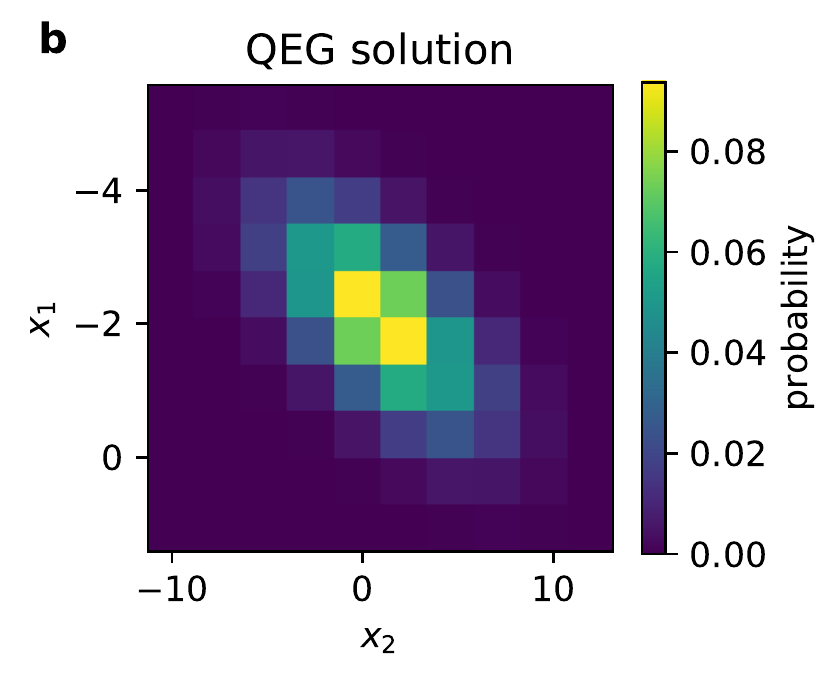}
    \caption{Synthetic probabilistic classifier. \textbf{a} 2-dimensional binning, with 10 bins per variable $x_1, x_2$; \textbf{b} QEG solution of corresponding maximum Shannon entropy probability distribution.}
    \label{fig:syntheticQEG}
\end{figure}

\newpage
\subsection*{Supplementary Note 2: Synthetic example for invariant groups detection}

We have generated a synthetic dataset of 10,000 samples and 5 columns, 4 of which represent the features $x_1\in[1, 2]$, $x_2\in[0.5, 1]$, $x_3\in[2, 5]$, $x_4\in[3, 7]$, and 1 represents the response $y$, governed by the function
\begin{equation}    y(\mathbf{x})=\sin\left(\ln\left(x_1^4x_2^3\right)\right)+\frac{\cos(5x_3)}{\sqrt{x_4}} + \sqrt{x_1^4x_2^3}.
\label{eq:synthetic}
\end{equation}
The synthetic function in Eq.~\ref{eq:synthetic} has the invariant group $x_1^4x_2^3$, which we aim at detecting. We have thus splitted the dataset into three parts: (i) a training set, (ii) a validation set to get insight of possible overfitting, (iii) a testing set to effectively evaluate the model performances. 

\begin{figure}
    \centering
    \includegraphics[width = 0.8\textwidth]{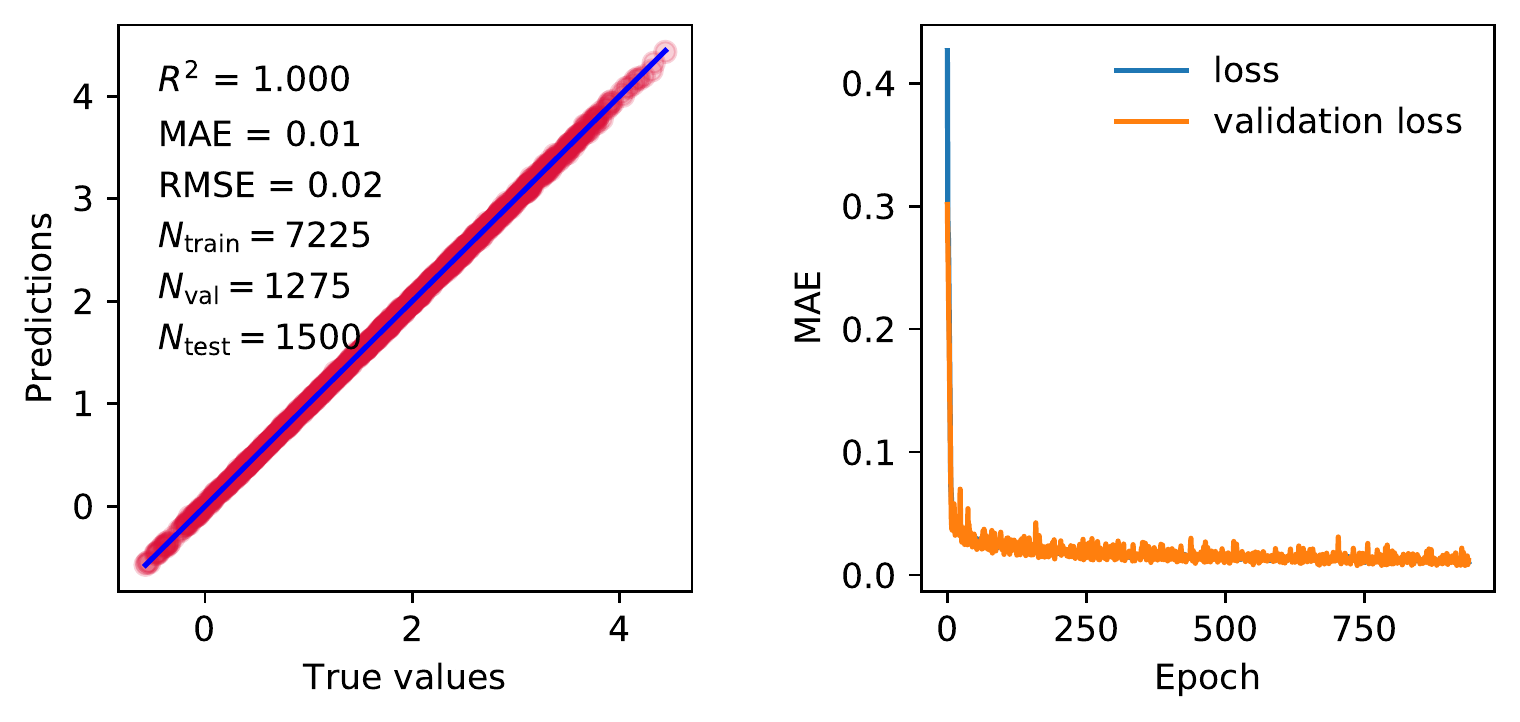}
    \caption{Predictions and corresponding loss curves for the DNN regression model. Model performances are shown in terms of coefficient of determination $R^2$, mean absolute error (MAE), and root mean squared error (RMSE), with the sizes of the training, the validation and the testing sets, $N_{\textrm{train}}$, $N_{\textrm{val}}$, $N_{\textrm{test}}$ respectively.}
    \label{fig:synthetic-invariant}
\end{figure}

Supplementary Fig.~\ref{fig:synthetic-invariant} shows the predictions over the testing set, together with the model performances and the corresponding loss with respect to the number of epochs. Specifically, no overfitting is found. More details about the DNN are shown in Supplementary Note 3. 

\begin{table*}
\centering
\caption{Invariant groups results for synthetic example.}
\begin{tabular}{crrrr}
\toprule
Features pair & mean $a$ & std.~dev. $a$ & mean $b$ & std.~dev. $b$\\
\midrule
$(x_1, x_2)$ & $\mathbf{0.799}$ & 0.047 & $\mathbf{0.596}$ & 0.060\\
$(x_1, x_3)$ & 0.954 & 0.034 & 0.054 & 0.294 \\
$(x_1, x_4)$ & 1.000 & 0.000 & -0.009 & 0.002\\
$(x_2, x_3)$ & 0.972 & 0.039 & 0.030 & 0.228\\
$(x_2, x_4)$ & 1.000 & 0.000 & -0.006 & 0.002\\
$(x_3, x_4)$ & 0.155 & 0.983 & -0.036 & 0.090\\

\bottomrule
\end{tabular}

\label{tab:invariantgroups}
\end{table*}
Supplementary table \ref{tab:invariantgroups} shows the means and the standard deviations of the coefficients $a$ and $b$ over 100 trials. The method identifies correctly that the pair $(x_1, x_2)$ has normalized exponents $a\approx0.8=4/\sqrt{3^2+4^2}$ and $b\approx0.6=3/\sqrt{3^2+4^2}$, with low standard deviations. For all the other cases, there are too high variances for at least one of the two exponents - e.g. $(x_1, x_3)$ - or there is too closeness to 0 for one of the two exponents - e.g. $(x_1, x_4)$, which cancels the group.

\newpage
\subsection*{Supplementary Note 3: Deep Neural Networks structures}
The structure of the Deep Neural Network for the superconductors example in the main text is the following:
\begin{itemize}
    \item optimizer Adam, learning rate 0.001;
    \item input layer 81 neurons;
    \item normalization layer adapted over the training set;
    \item dense layer, 128 neurons, activation ReLU;
    \item dense layer, 64 neurons, activation ReLU;
    \item dense layer, 64 neurons, activation ReLU;
    \item dense layer, 32 neurons, activation ReLU;
    \item dense layer, 32 neurons, activation ReLU;
    \item dense layer, 16 neurons, activation ReLU;
    \item dense layer, 16 neurons, activation ReLU;
    \item dense layer, 8 neurons, activation ReLU;
    \item dense layer, 1 neuron, activation linear;
    
\end{itemize}

The structure of the Deep Neural Network for the synthetic example in the Supplementary Note 2 is the following:
\begin{itemize}
    \item optimizer Adam, learning rate 0.001;
    \item input layer 4 neurons;
    \item normalization layer adapted over the training set;
    \item dense layer, 64 neurons, activation Leaky ReLU;
    \item dense layer, 64 neurons, activation Leaky ReLU;
    \item dense layer, 32 neurons, activation Leaky ReLU;
    \item dense layer, 32 neurons, activation Leaky ReLU;
    \item dense layer, 16 neurons, activation Leaky ReLU;
    \item dense layer, 16 neurons, activation Leaky ReLU;
    \item dense layer, 8 neurons, activation Leaky ReLU;
    \item dense layer, 1 neuron, activation linear;
    
\end{itemize}

In particular, in both models we have employed the early stopping regularization, with $\textrm{patience} = 200$.


\newpage
\subsection*{Supplementary Note 4: ETR/ETC hyperparameter tuning}

The pipeline generating ETR, ETC-vanilla and ETC-SMOTE models is given by the steps in the following: (i) drop of features with linear correlation $> 0.9$, (ii) drop of features with no variance, (iii) oversampling with SMOTE - only for ETC-SMOTE, (iv) drop of the least relevant features in terms of the \url{f_regression} test [1] - for ETR - or \url{f_classif} test [1] - for ETCs, (v) training of the model. 

The space of hyperparameters we have explored for the ETR-based pipeline is composed by:

\begin{itemize}
    \item the percentage of the most relevant features to retain according to the \url{f_regression}/\url{f_classif} test, at point (iv), to be chosen among [50, 75, 100];
    \item the number of estimators in the ETR/ETC model, to be chosen among [100, 250, 500, 750, 1000];
    \item the number of max features, i.e., the number of features to keep when looking for the best split, to be chosen among [1, 0.9, 0.8, 0.7, 0.6, 0.5].
\end{itemize}

In the aforementioned space, we have performed the hyperparameter tuning in 5-fold cross validation (for the ETR) or in stratified 5-fold cross validation (for ETCs) by means of \url{GridSearchCV} [1], thus exploring all the possible combinations, ensuring the absence of data leakage and preventing overfitting.


\newpage
\subsection*{Supplementary Note 5: SHAP}
With the Tree SHAP algorithm, which is tailored for tree-based models like ETR [2,3], we identify the most crucial features during model training and validation, quantifying the influence of each feature on the output. This approach is based on the traditional Shapley value, which originally found use in game theory. There, the issue of allocating a proportionate reward to each player in a cooperative game is handled based on the actual contribution provided to the coalition's common goal.

The relevance of the $i\rm{-th}$ descriptor in a model is determined by comparing the model $f_{S\cup\{i\}}(\mathbf{x}_{S\cup\{i\}})$ trained with that explanatory variable to another model $f_{S}(\mathbf{x}_S)$ trained without that feature, where $\mathbf{x}_{S\cup\{i\}}$ and $\mathbf{x}_S$ represent, respectively, the values of the input feature over the subsets $S\cup\{i\}$ and $S$. The difference between the predictions  $f_{S\cup\{i\}}(\mathbf{x}_{S\cup\{i\}}) - f_{S\cup\{i\}}(\mathbf{x}_S)$ is thus computed. The relevance value of the $i\textrm{-th}$ descriptor is represented by this difference weighted over all possible subsets $S$. The importance of the $i\textrm{-th}$ turns out to be
\begin{equation}
    \phi_i = \sum_{S\subseteq F\setminus \{i\}}\frac{|S|!(|F|-|S|-1)!}{|F|!}\left(f_{S\cup\{i\}}(\mathbf{x}_{S\cup\{ i\}})-f_S(\mathbf{x}_S)\right)
\end{equation}
where $|\cdot|$ denotes the number of elements.


\newpage
\subsection*{Supplementary Note 6: Outliers removal}
During the dataset creation, as the last preprocessing step we have removed the top four materials - in terms of $T_{\rm{c}}$ - as outliers. This is justified by binning the 7230 (before the removal) critical temperatures according to the Sturges' rule, by which the optimal number of bins is $1 + \textrm{ceil}(\log_2(h))$ [4], where $h=7230$. We end up with the binning shown in Supplementary Fig.~\ref{fig:binning}. The top four $T_{\rm{c}}\rm{s}$ are indeed far from the distribution of the other materials, so we dropped them.

\begin{figure}[h]
    \centering
    \includegraphics[width = 0.8\textwidth]{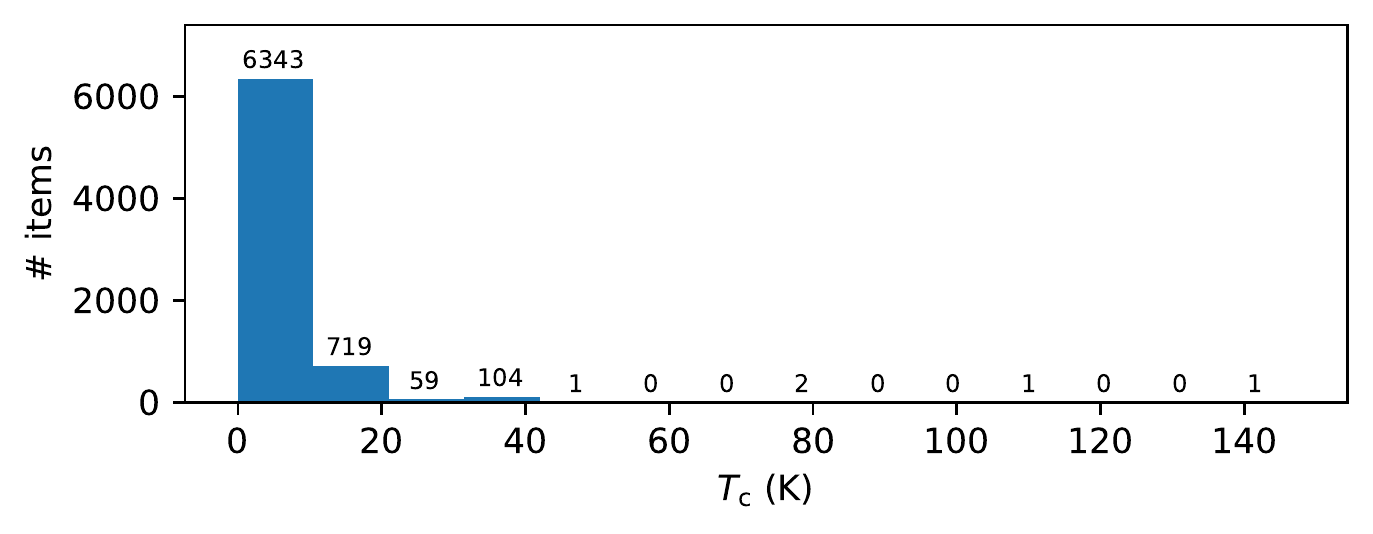}
    \caption{Binning over 7230 critical temperatures of the dataset.}
    \label{fig:binning}
\end{figure}


\newpage
\subsection*{Supplementary Note 7: Extra Trees}
In random forests, a number of decision trees is produced, over random partitions of the input space. On each one of them, a decision tree is constructed; their predictions are aggregated by taking the mean. Specifically, during the building of a tree, the best split is determined either from all input features or from a random subset of size \url{max_features}. 

On the contrary, on Extra Trees-based models (ETR and ETC), only a random subsets of possible splits is explored, and the best one of them is picked as the splitting rule [1].


\newpage
\subsection*{Supplementary Note 8: ROC curves}

\begin{figure}[h]
    \centering
    \includegraphics[width = .6\textwidth]{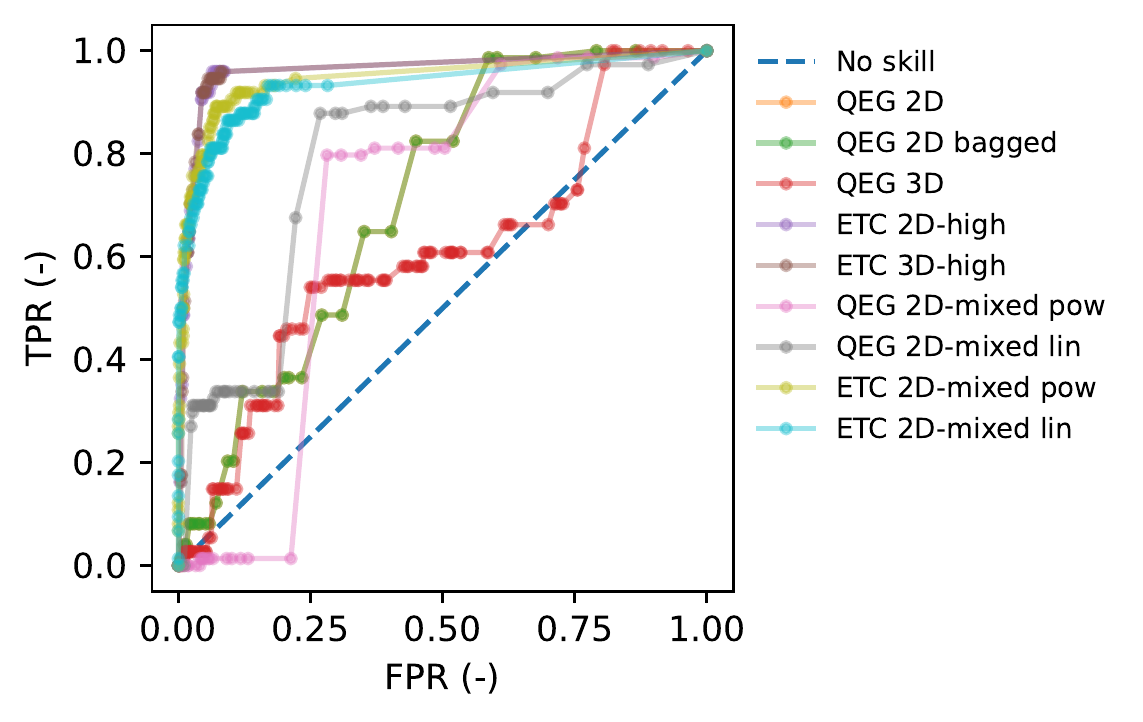}
    \includegraphics[width = .6\textwidth]{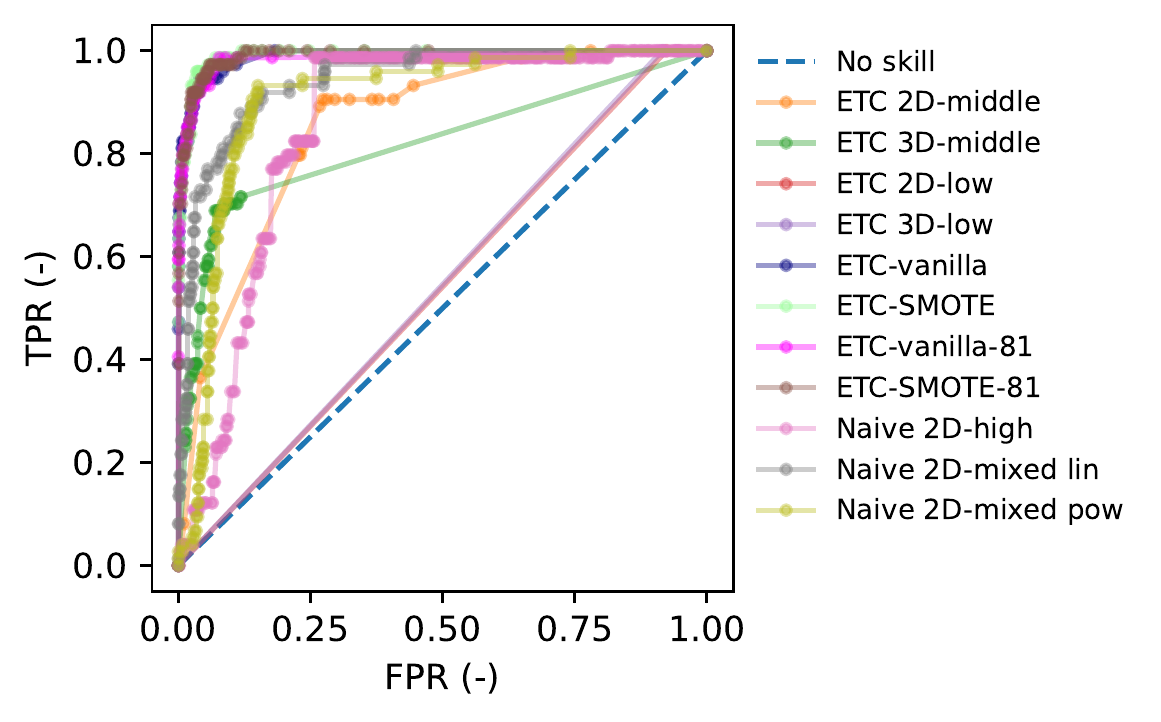}
    \caption{ROC curves for the \emph{No skill}, QEG 2D, QEG 2D bagged, QEG 3D, ETC 2D-high, ETC 3D-high, QEG 2D-mixed pow, QEG 2D-mixed lin, ETC 2D-mixed lin, ETC 2D-mixed pow, ETC-2D-middle, ETC 3D-middle, ETC 2D-low, ETC 3D-low, ETC-vanilla, ETC-SMOTE, ETC-vanilla-81, ETC-SMOTE-81 classifiers, with discriminating threshold of $T_{\rm{c}}=15\, \si{\kelvin}$.}
    \label{fig:ROC1}
\end{figure}


\newpage
\subsection*{Supplementary Note 9: Optimization for mixed features}

\textbf{Multi-objective optimization.} For finding the mixed features $x^{\rm{pow, least, 35}}$, $x_{1,2}^{\rm{pow, uto, 35}}$, $x_{1,2}^{\rm{pow, uto, 15}}$, $x_{1,2}^{\rm{lin, uto, 15}}$, $x_{1, 2}^{\rm{pow, 3class}}$ we have employed a multi-objective optimization by means of a Pareto front in MATLAB. 
Here we have considered only the Utopia point - namely, the closest to the origin (also for the 3 classes example) - and the \emph{least} point - namely, the one assessing the lowest overlap between the two classes, as prescribed by the Bhattacharyya coefficient $BC(P, Q) = \sum_{x\in\mathcal{X}}\sqrt{P(x)Q(x)}$, with $P$, $Q$ being two discrete probability distributions defined on the same domain $\mathcal{X}$. This is equivalent to maximize the Bhattacharyya distance $-\ln{BC(P, Q)}$    [5,6]. 

\textbf{Single objective optimization.} For finding the mixed features $x_{1, 2}^{\rm{lin, 25}}$ we have employed a single objective optimization by means of a Genetic Algorithm and a Pattern Search in MATLAB. In particular, we have used the Genetic Algorithm to construct a first tentative solution, improved by means of the MATLAB \texttt{patternsearch} routine to get the solution minimizing the number of negative neighbors around a positive sample in a given radius. Thus, here we do not consider the distance between the two distributions.



\newpage
\subsection*{Supplementary Note 10: Naive Gaussian Bayesian Classifier}

In Naive Bayes Classifiers, the Bayes theorem, with the \emph{naive} assumption of conditional independence between any pair of features, is applied. Given the class variable $y$ and a set of features $x_1, \dots, x_n$, the Bayes theorem yields 
\begin{equation}
    P(y|x_1, \dots, x_n) = \frac{P(y)P(x_1, \dots, x_n|y)}{P(x_1, \dots, x_n)}
\end{equation}
that, with the naive independence condition $P(x_i|y, x_1, \dots, x_{i-1}, x_{i+1}, \dots x_n) = P(x_i|y)$, becomes
\begin{equation}
    P(y|x_1, \dots, x_n) = \frac{P(y)\prod_{i=1}^nP(x_i|y)}{P(x_1, \dots, x_n)}.
\end{equation}
Since $P(x_1, \dots, x_n)$ is constant given the input, the classification rule can be written as
\begin{equation}
    \hat{y} = \arg\max_yP(y)\prod_{i=1}^nP(x_i|y).
\end{equation}
In the Naive Gaussian Classifier employed in this paper, $P(x_i|y)$ is the Gaussian
\begin{equation}
    P(x_i|y) = \frac{1}{\sqrt{2\pi\sigma_y^2}}\exp{\left(-\frac{(x_i-\mu_y)^2}{2\sigma_y^2}\right)},
\end{equation}
where $\mu_y$ and $\sigma_y$ are estimated by means of maximum likelihood.

For further details, please refer to refs.~[7,1]. 

In this work, we have employed all the default hyperparameters for Gaussian Naive Classifiers, as implemented in Scikit-learn [1].

\newpage
\subsection*{Supplementary Note 11: Naive Gaussian Bayesian Classifier - example at $T_{\rm{c}} = 35\, \si{\kelvin}$}

\begin{figure}
    \centering
    \includegraphics[width = 0.7\textwidth]{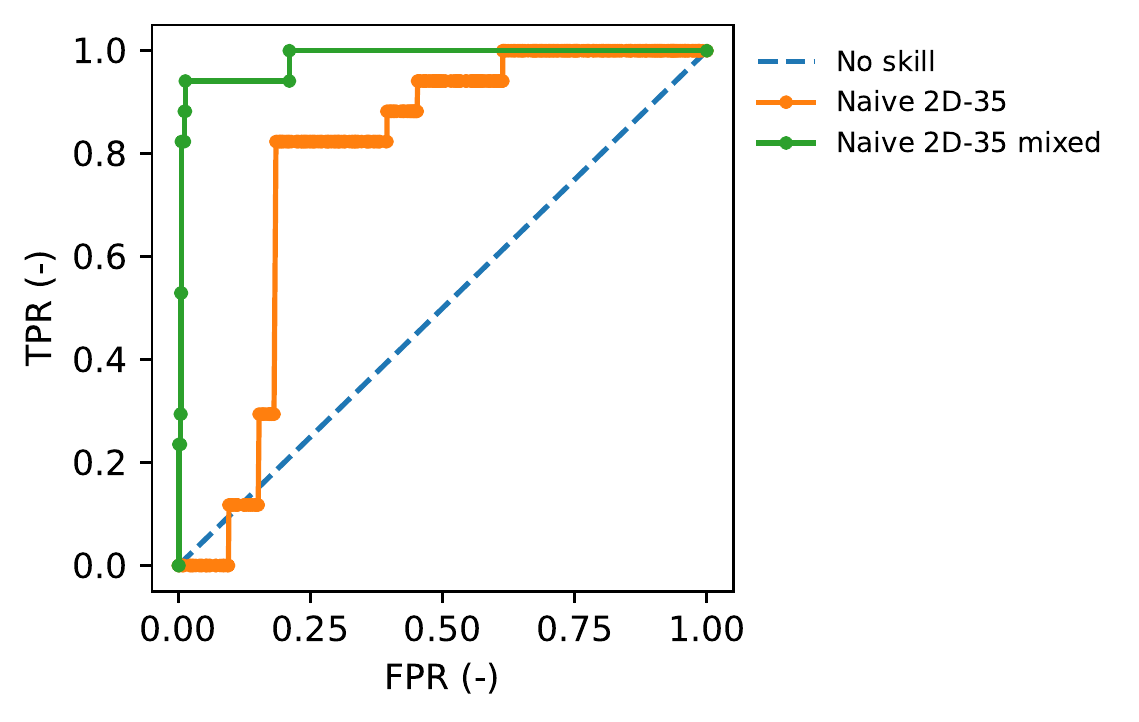}
    \caption{ROC curves for the \emph{No skill}, Naive 2D-35, Naive 2D-35 mixed classifiers, with discriminating threshold of $T_{\rm{c}} = 35\, \si{\kelvin}$.}
    \label{fig:bayes35}
\end{figure}

In the main text, we have given performances of Naive Gaussian Bayesian classifiers only for the threshold of $T_{\rm{c}} = 15\, \si{\kelvin}$. Here we analyze the case of a threshold $T_{\rm{c}} = 35\, \si{\kelvin}$, comparing two classifiers, namely the former trained with the first two features of the SHAP ranking, the latter trained with the two optimized features $x_1^{\rm{pow, uto, 35}}$, $x_2^{\rm{pow, uto, 35}}$ (already used in the main text example of Fig.~8). Supplementary table \ref{tab:performances} shows a sharp improvement in terms of all the metrics if employing the mixed features (Naive 2D-35 mixed) with respect to the SHAP features (Naive 2D-35); Supplementary Fig.~\ref{fig:bayes35} shows the corresponding ROC curves.

\begin{table*}
\centering
\caption{Performances of the trained classifiers.}
\label{tab:performances}
\begin{tabular}{lrrrrr}
\toprule
& AUC & $\xi_{J, \rm{max}}$ & $J_{\rm{max}}$ & $\xi_{F_1,\rm{max}}$ &  $F_{1, \textrm{max}}$\\
\midrule
No skill & 0.50 & - & - & - & - \\
Naive 2D-35 & 0.78 & 0.021 & 0.64 & 0.021 & 0.12 \\
Naive 2D-35 mixed & 0.98 & 0.526 & 0.93 & 0.532 & 0.76 \\
\bottomrule
\end{tabular}
\end{table*}

\clearpage
\section*{References}

\begin{enumerate}[label={[\arabic*]}]
\item Pedregosa, F. et al. Scikit-learn: Machine learning in Python. Journal of Machine Learning
Research 12, 2825–2830 (2011).
\item Lundberg, S. M. et al. From local explanations to global understanding with explainable ai for
trees. Nature machine intelligence 2, 56–67 (2020).
\item Lundberg, S. M. \& Lee, S.-I. A unified approach to interpreting model predictions. Advances in
neural information processing systems 30 (2017).
\item Sturges, H. A. The choice of a class interval. Journal of the american statistical association 21,
65–66 (1926).
\item Bhattacharyya, A. On a measure of divergence between two statistical populations defined by
their probability distributions. Bull. Calcutta Math. Soc. 35, 99–109 (1943).
\item Bhattacharyya, A. On a measure of divergence between two multinomial populations. Sankhy¯a:
the indian journal of statistics 401–406 (1946).
\item Zhang, H. The optimality of naive bayes. Aa 1, 3 (2004).

\end{enumerate}